\documentclass[twocolumn,english,prb,showpacs]{revtex4}

\usepackage[T1]{fontenc}
\usepackage[latin9]{inputenc}
\usepackage{float}
\usepackage{amsmath}
\usepackage{graphicx}
\usepackage{amssymb}
\usepackage{esint}
\usepackage{color}
\usepackage{babel}

\begin{document}

\title{Transmission phase shifts of Kondo impurities}

\author{Assaf Carmi$^{1, 2}$, Yuval Oreg$^1$, Micha Berkooz$^2$, and David Goldhaber-Gordon$^{1,3}$\\
\textit{$^1$Department of Condensed Matter Physics, Weizmann Institute
of Science, Rehovot, 76100, Israel\\$^2$Department of Particle Physics and Astrophysics, Weizmann Institute
of Science, Rehovot, 76100, Israel\\$^3$Department of Physics, Stanford University, Stanford, California, 94305, USA}}
\begin{abstract}
We study the coherent properties of transmission through Kondo impurities, by considering an open Aharonov-Bohm ring with an embedded quantum dot. We develop a novel many-body scattering theory which enables us to calculate the conductance through the dot $G_d$, the \emph{transmission} phase shift $\varphi_t$, and the normalized visibility $\eta$, in terms of the single-particle $\mathcal{T}$-matrix.
For the single-channel Kondo effect, we find at temperatures much below the Kondo temperature $T_K$ that $\varphi_t=\pi/2$ without any corrections up to order $(T/T_K)^2$. The visibility has the form $\eta=1-(\pi T/T_K)^2$.
For the non-Fermi liquid fixed point of the two channel Kondo, we find that $\varphi_t=\pi/2$ despite the fact that a scattering phase shift is not defined. The visibility is $\eta=1/2(1+4\lambda\sqrt{\pi T})$ with $\lambda\sim 1/\sqrt{T_K}$, thus at zero temperature exactly half of the conductance is carried by single-particle processes, and coherent transmission may actually increase with temperature.
We explain that the spin summation masks the inherent scattering phases of the dot, which can be accessed only via a spin-resolved experiment. In addition, we calculate the effect of magnetic field and channel anisotropy, and generalize to the k-channel Kondo case.
\end{abstract}

\pacs{71.27.+a,72.15.Qm,75.20.Hr}

\maketitle

\section{Introduction}

%A quantum dot that is attached to electronic leads may exhibit many interesting physical %effects. When the dot is capacitively coupled to a gate electrode, the number of %electrons in the dot is controlled by the gate voltage.
%
If the ground state of a quantum dot has a fixed number of electrons, decreasing the temperature to below the charging energy of the dot reduces the conductance $G_d$ through the dot because of Coulomb blockade~\cite{Blockade1,RevModPhys.64.849,0268-1242-11-3-003,Blockade4}. If the electron occupancy is odd, lowering the temperature even further increases $G_d$, until it reaches (for a symmetrically-coupled dot) $2e^2/h$ at zero temperature~\cite{GlazmanRaikh,nature.391.156,science.281.5376.540}.

The enhancement of the conductance is due to the single-channel Kondo (1CK) effect~\cite{prog.theor.phys.32.37}, in which the dot acts as a magnetic impurity that interacts with the spins of the electrons in the surrounding leads. At low temperature, below a characteristic temperature $T_K$, a spin resonance is formed, and the conductance through the resonance is perfect and equals $e^2/h$ per spin.
The physics of 1CK at low energy can be described by a Fermi liquid theory: at zero temperature, all the particles that scatter off the impurity are scattered into single-particle states, where the incoming and outgoing states are connected by a $\pi/2$ scattering phase shift~\cite{J.low.temp.phys.17.31} (see also Sec.~\ref{sec: s-matrix discussion}).

The 1CK physics can be generalized to more complex models, known as multi-channel Kondo, where a few independent channels compete to screen the impurity~\cite{J.physique(paris).41.193}. In the two-channel Kondo (2CK) case, when the couplings of the two channels to the impurity are identical the system flows to a non-Fermi liquid fixed point at zero temperature. At a non-Fermi liquid fixed point, the simple picture of elastic scattering of single particles is no longer valid. At zero temperature, a single particle that is scattered off a 2CK impurity can be scattered only into a many-body state~\cite{PhysRevB.48.7297,PhysRevB.46.10812}. Thus, there is no elastic single-particle scattering off a 2CK impurity at the non-Fermi liquid fixed point.
The 2CK system was first discussed as a purely theoretical problem~\cite{J.physique(paris).41.193}, but it was soon invoked as a candidate explanation for remarkable low-energy properties of some heavy fermion
materials~\cite{PhysRevLett.59.1240,Besnus1988471,PhysRevLett.67.2882,PhysRevB.79.012411} and glassy metals~\cite{PhysRevLett.45.211,PhysRevLett.69.2118,PhysRevLett.72.1064,PhysRevLett.94.236603,1742-6596-200-1-012021} and more recently in graphene~\cite{PhysRevB.77.045417,Souza2009,1742-5468-2010-01-P01007,0295-5075-90-6-67001}.
In the past decade, a few single-impurity realizations of the 2CK system were proposed~\cite{PhysRevB.51.1743,PhysRevB.51.17827,PhysRevB.68.041311,PhysRevLett.90.136602,J.phys.condens.Matter.16.749},
offering the hope of microscopically manipulating system parameters, and one of the proposals~\cite{PhysRevLett.90.136602} was built and measured~\cite{nature.446.167}. The conductance through a 2CK impurity, within one of the two channels, at the non-Fermi liquid fixed point is $e^2/2h$ per spin, assuming equal coupling to two leads in that channel~\cite{PhysRevB.48.7297}.

Given that there are no elastic single-particle scattering events off the impurity in the non-Fermi liquid fixed point, one might imagine that the transport through a 2CK impurity has no coherent part. In this work, we show that at this fixed point exactly half of the conductance is carried by coherent processes~\cite{footnote_abstract}. This is because in a transport measurement through a single-level quantum dot there are (at least) two leads that are attached by tunneling to the dot. The electrons that interact with the effective spin of the dot are described by an operator $\psi$, a linear combination of electron operators in the two leads. Another linear combination of electron operators in the two leads, $\xi$, is decoupled from the dot. While there are no elastic single $\psi$-particle scattering events, coherent transport via $\xi$-particles is possible.

\begin{figure}
\includegraphics[bb=25bp 281bp 550bp 530bp,clip,scale=0.4]{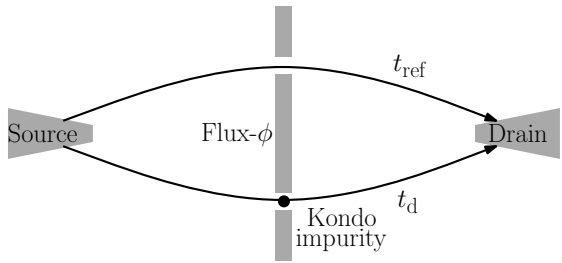}
\caption{Schematic picture of a two-path setup. Electrons are sent from the source lead toward the drain lead through two paths, whose partial waves which interfere with each other. The transmission amplitudes of the two paths are $t_{\text{d}}$ and $t_{\text{ref}}$, and they encircle a magnetic flux $\phi$. The coherent transport through the Kondo impurity can be studied by embedding it into one of the paths.}
\label{fig:Schematic double slit}
\end{figure}

The coherent properties of the transport through an impurity can be measured in a two-path experiment, in which electrons are sent from a source lead through two possible paths to a drain lead (see Fig.~\ref{fig:Schematic double slit}). We assume that the propagations along the different paths are independent of each other, namely, changes in the properties of one path do not affect the propagation along the other path.
One of the paths contains the impurity of interest, and the two paths encircle a magnetic flux $\phi$. The interference between the two paths depends on $\phi$ through the Aharonov-Bohm (AB) effect. Hence, the conductance of the setup contains two parts: a flux-independent part, which is related to the separate conductances of the two paths, and a flux-dependent part, which is related to the interference of the two paths.

Two measurable quantities can be extracted from a two-path experiment: the transmission phase shift of the flux-dependent conductance, and the ratio between the amplitude of the flux-dependent part and the flux-independent part of the conductance. We cast the source-to-drain conductance of the two-path device into the form
\begin{equation}
G_{\rm sd}=G_{\text{d}}+G_{\text{ref}}+2\sqrt{\eta}\sqrt{G_{\text{d}}G_{\text{ref}}}\cos\left(\frac{e\phi}{\hbar c}+\varphi_t\right),\label{eq: definition of eta and phi}
\end{equation}
where $G_{\text{d}}$ is the conductance through the path with the impurity when the reference path is switched off, and $G_{\text{ref}}$ is the conductance through the reference path when the impurity's conduction
is switched off. The impurity will generally be realized as one or more quantum dots, so we will interchangeably refer to ``impurity'' and ``dot'' depending on context. We assume that the paths are independent: manipulations of the dot (for example, with gate potential) do not influence the conductance of the reference path, and vice versa.
The \emph{transmission phase}, $\varphi_t$, is related to the relative phase between the two paths, and the \emph{normalized visibility} $\eta$ is related to the size of the coherent part of the conductance compared to the total conductance. Note: the phase of the reference path is arbitrary, determined by path length, potential landscape, etc. So is the phase of the path with the impurity, excluding the transmission phase of the impurity itself. Below we assume for simplicity that each of these phases is $0$ mod $2\pi$, so that $\varphi_t$ is purely the transmission phase of the impurity itself. The definition of $\eta$, implicit in Eq.~\eqref{eq: definition of eta and phi}, is such that for Fermi liquids, at zero temperature and without spin, $\eta=1$. This can be easily checked by applying the Landauer formalism~\cite{5392683,PhysRevB.31.6207,PhysRevLett.57.1761,imry2002introduction} for the two-path experiment setup.

The normalized visibility $\eta$ can be reduced to below one by four mechanisms: First, if the transmitted electrons accumulate an energy-dependent phase when they are scattered through the impurity, or just along either path, then at nonzero temperature $\eta$ is reduced because of the thermal averaging. Second, if the phase depends on the spin, the spin summation can also reduce $\eta$. Third, if part of the conductance is carried by incoherent scattering processes, where single electrons are scattered into many-body states, the interference and therefore $\eta$ are reduced. Fourth, electrons that are subjected to external dephasing lose their coherence, so external dephasing also decreases the interference and $\eta$. External dephasing depends on the specific model and the details of the setup. Hence, we focus mainly on the first three mechanisms, and only qualitatively explore the effect of external dephasing on $\eta$.

Since $G_{\text{d}}$ and $G_{\text{ref}}$ can be measured directly, the normalized visibility can be experimentally determined. This requires two measurements: the conductance through one of the paths, and the two-path conductance. Measuring the transmission phase of a 1CK impurity in a two-path setup was already suggested before~\cite{PhysRevLett.84.3710}, and the predicted $\varphi_t=\pi/2$ was measured~\cite{PhysRevLett.100.226601}, demonstrating coherent electron transmission through a many-body state. Yet, no special attention was given to the amplitude of the flux-dependent part of the conductance. In particular, non-Fermi liquid cases, where $\eta$ can give information on the underlying physics (and also $\varphi_t$ is different from that in the 1CK case), were not treated.

In Sec.~\ref{sec: s-matrix discussion}, we relate $\varphi_t$ and $\eta$ to the single-particle elements of the $\mathcal{T}$-matrix, $\mathcal{T}_{\psi\psi}$ (where the $\psi$-particles are the particles that interact with the dot). Using arguments of many-body scattering, we find a relation between the coherent and the incoherent parts of the conductance $G_{\text d}$, and rederive the known expression for the conductance~\cite{PhysRevB.64.045328,springerlink:10.1007/978-3-540-72632-6_2,review_DO}
\begin{equation}
G_{\text d}=\mathcal{G}_0\int d\epsilon\left(-\frac{\partial f}{\partial\epsilon}\right)2Im\left\{\mathcal{T}_{\psi\psi}\right\},
\end{equation}
where $\mathcal{G}_0$ is the quantum conductance multiplied by a symmetry factor related to relative coupling to different leads. We also derive the following relations
\begin{equation}
\varphi_t=\arg\left(\langle\mathcal{T}_{\psi\psi}\rangle\right), \; \; \; \; \; \; \; \; \;
\eta=\frac{\left|\langle\mathcal{T}_{\psi\psi}\rangle\right|^2}{2Im\langle\mathcal{T}_{\psi\psi}\rangle},
\end{equation}
where $\langle\mathcal{T}_{\psi\psi}\rangle=\frac{1}{2}\sum_s d\epsilon\left(-\frac{\partial f}{\partial\epsilon}\right)\mathcal{T}_{s,\psi\psi}$ is the thermal- and spin-averaged value of the $\mathcal{T}$-matrix. Expressions for the dephasing rate and the ratio between the inelastic scattering cross section and the total cross section, both related to the normalized visibility $\eta$, appear in the literature~\cite{PhysRevLett.93.107204,PhysRevLett.96.226601,PhysRevB.75.235112,Zarand20075}.
We show that spin summation has a crucial effect on $\varphi_t$ and $\eta$ of Kondo impurities. Up to second order in $T/T_K$ (and $B/T_K$), spin summation locks the value of $\varphi_t$ at $\pi/2$ independent of the actual phases that electrons accumulate when they cross the dot. Moreover, spin summation reduces $\eta$ significantly, even when all the conductance is carried by coherent single-particle scattering. The $\pi/2$ phase-lock, and reduction of $\eta$, can be avoided if one measures the conductance of each spin separately, and extracts directly the transmission phase of each spin, $\varphi_{ts}$, separately. A concrete realization of Kondo impurities in quantum dots, with access to each spin separately, was proposed by some of the present authors in Ref.~\onlinecite{PhysRevLett.106.106401}.

The main results of this work are as follows:\newline
It is known that in the 1CK case, at zero temperature, the transmission phase equals the scattering phase shift of the 1CK~\cite{PhysRevLett.84.3710}, $\varphi_t=\pi/2$. Since all the electrons are elastically scattered, the normalized visibility is $\eta=1$. However, when a magnetic field is applied, regular non-spin-resolved measurements of the conductance miss the magnetic field corrections. In this case, we find that the transmission phase $\varphi_t$ remains $\pi/2$ to second order in $B$, even though the scattering phase shift for each spin depends on the magnetic field~\cite{J.low.temp.phys.17.31,J.physique(paris).41.193},
$\delta_{\psi,s}=\alpha_s(\pi/2-B/T_K)$ ($\alpha_\uparrow=1$ and $\alpha_\downarrow=-1$). In order to reveal the magnetic field dependence of the phase shift, one needs to perform a spin-resolved measurement, namely, to measure the conductance of each spin separately.

In the non-Fermi liquid fixed point of the 2CK case, we find that at zero temperature $\eta=1/2$, that is, exactly half of the conductance is carried by single-particle transmissions~\cite{PhysRevLett.93.107204}. The electrons that elastically transmit through the 2CK impurity accumulate a $\varphi_t=\pi/2$ phase when they cross the impurity. In the presence of a finite magnetic field, the system flows under renormalization group to a Fermi liquid fixed point, at zero temperature,
rather than the non-Fermi liquid one~\cite{PhysRevB.45.7918}. At this Fermi liquid fixed point, we find that again, one needs to perform a spin-resolved measurement. A spin-summed measurement will lead, at zero temperature, to a transmission phase $\varphi_t=\pi/2$ and a normalized visibility $\eta=1/2$, despite the fact that the spin-dependent scattering phase shifts are $\delta_{\psi,s}=\alpha_s\pi/4$, and despite the fact that the conductance is carried exclusively by single-particle scattering [see Eqs.~\eqref{eq: phase lock pi/2} and ~\eqref{eq: normalized visibility two spins}]. Measurement of each spin separately, however, will lead to the desired $\eta=1$, and $\varphi_t=\alpha_s\pi/4$.

The rest of the paper is organized as follows: in Sec.~\ref{sec: transmission phase and visibility}, we briefly review possible realizations of electronic two-path experiments, and discuss what we learn from their analysis. We define the two measurable quantities, $\varphi_t$ and $\eta$, and discuss their physical meaning. In Sec.~\ref{sec: s-matrix discussion}, we develop a scattering approach to the transport through an impurity, similar to the Landauer formalism\cite{5392683,PhysRevB.31.6207,PhysRevLett.57.1761,imry2002introduction} for the non-interacting case. We consider a many-body scattering matrix to include both elastic single-particle scattering and inelastic single-particle to multi-particle scattering. We rederive the conductance through the impurity and give the mathematical expressions for $\varphi_t$ and $\eta$. In Sec.~\ref{sec:results}, we focus on Kondo impurities, and give the results for $\varphi_t$ and $\eta$ for several Kondo fixed points. We also briefly discuss the influence of possible external dephasing on the normalized visibility. Finally, we summarize our results and conclusions in Sec.~\ref{conclusions}. In Appendix \ref{s-matrix appendix}, we give a detailed derivation of the multi-particle scattering approach for the conductance, transmission phase, and normalized visibility. In Appendix \ref{two paths setup appendix}, we give more details about a possible two-path setup that can be tuned to fulfill the theoretical assumptions we have made in our analysis.

\section{Two-path experiments, transmission phase, visibility, and normalized visibility}\label{sec: transmission phase and visibility}
In this section, we discuss two-path setups and define the transmission phase $\varphi_t$ and the normalized visibility $\eta$. We emphasize that the normalized visibility, $\eta$, is distinct from the more common definition of the visibility.

The prototype of two-path experiments is the double-slit experiment. In a double-slit experiment particles are launched toward the double slit, where they split into partial waves which interfere with each other. In the electronic version of the double-slit experiment, schematically drawn in Fig. \ref{fig:Schematic double slit}, a coherent electron beam is emitted from a source lead toward a drain lead, via a beam splitter that allows electron flow along two different paths that encircle a magnetic flux $\phi$. The source-to-drain conductance is given by
\begin{equation}
G_{\rm sd}=\frac{e}{h}\sum_s\int d\epsilon \left(-\frac{\partial f}{\partial\epsilon}\right)T_s(\epsilon),\label{eq: G_sd}
\end{equation}
where $T_s(\epsilon)$ is the probability for an incoming electron with energy $\epsilon$ and spin $s$ to be transmitted through the double slit, and $f(\epsilon)$ is the Fermi-Dirac distribution function. If all the electrons that pass through the double slit do so elastically and coherently, the probability $T_s(\epsilon)$ is given by~\cite{5392683}
\begin{align}
T_s=&\left|t_{\text{d},s}\right|^2+\left|t_{\text{ref},s}\right|^2+2\left|t_{\text{d},s}t_{\text{ref},s}\right|\cos\left(\frac{e\phi}{\hbar c}+\theta_s\right),\label{eq: transmission probability}
\end{align}
where $t_{\text{d},s}$ and $t_{\text{ref},s}$ are the transmission amplitudes of the two slits. The transmission amplitudes are complex quantities with a phase difference, $\frac{e\phi}{\hbar c}+\theta_s$, between them. The phase difference contains a contribution $\theta_s$ determined by the details of the transmission through the double-slit setup, and a magnetic-flux-dependent part $\frac{e\phi}{\hbar c}$ coming from the AB effect.

Equation (\ref{eq: transmission probability}) is valid only if all the electrons are coherently transferred through the double slit~\cite{PhysRevB.65.045316}. If some of the electrons are transferred incoherently through one of the slits, then, since these electrons do not interfere, the flux-dependent term of $T_s$ is reduced.
If we embed into one of the paths a quantum dot (as in the lower path in Fig. \ref{fig:Schematic double slit}), we can examine the dot's coherence properties by measuring the conductance. In such a device, the phase that electrons accumulate as they cross the dot is encoded in the relative phase between the two paths $\theta_s$.

In experiments, the measured source to drain conductance is typically cast in the form
\begin{equation}
G_{\rm sd}=G_0+G_\phi\cos\left(\frac{e\phi}{\hbar c}+\varphi_t\right).\label{eq: typical conductivity}
\end{equation}
$G_0$ is the part of the conductance which is independent of the magnetic flux, and is related to the independent conductances of the two paths, and $G_\phi$ is the amplitude of the flux-dependent part of the conductance. In the general case, $\varphi_t$ is different from $\theta_s$, but if $t_{\text{d},s}$, $t_{\text{ref},s}$, and $\theta_s$ are independent of spin and energy, then $\varphi_t=\theta_\uparrow=\theta_\downarrow$. In standard two-path experiments, the ratio ${G_\phi}/G_0$, is called "visibility", and it measures the strength of the flux-dependent conductance oscillation compared to the average conductance.

The ratio ${G_\phi}/G_0$ can be reduced by several mechanisms. Trivially, a mismatch between the transmission amplitudes, $|t_{\text{d}}|\neq |t_{\text{ref}}|$, decreases the ratio $|t_{\text{d}}t_{\text{ref}}|/(|t_{\text{d}}|^2+|t_{\text{ref}}|^2)$, and therefore reduces ${G_\phi}/G_0$. In addition to the trivial transmission amplitude mismatch, four other mechanisms noted earlier can reduce ${G_\phi}/G_0$: thermal averaging, spin averaging, inelastic scattering, and externally-induced dephasing.

There is a conceptual difference between transmission amplitude mismatch of the two paths, and the other three mechanisms for ${G_\phi}/G_0$ reduction (we assume for the moment that there is no external dephasing). Unlike the transmission amplitude mismatch, these other mechanisms cannot be probed by simple single-path conductance measurements of the system. To isolate the transmission mismatch from elastic versus inelastic scattering and energy or spin dependent phase, we decompose the conductance (\ref{eq: typical conductivity}) into the form of Eq.~\eqref{eq: definition of eta and phi}:
\begin{equation}
G_{\text{d}}+G_{\text{ref}}+2\sqrt{\eta}\sqrt{G_{\text{d}}G_{\text{ref}}}\cos\left(\frac{e\phi}{\hbar c}+\varphi_t\right)\; .\nonumber
\end{equation}
$G_{\text{d}}$ and $G_{\text{ref}}$ are the independent conductances through the two paths, which can be measured directly by closing off one and then the other path. Equation (\ref{eq: definition of eta and phi}) defines a new quantity, the \emph{normalized visibility} $\eta$. If all the electrons transmit coherently through the two paths, and accumulate the same phase, then $\eta=1$, independent of possible transmission amplitudes mismatch.

We want to make a comment about the feasibility of interference measurements in two-path experiments: In real experiments, there is a typical coherence length, $l_{\rm coh}$, along which the propagating electrons preserve their coherence. This length depends on the details of the realization of the two-path setup, and we assume that it is much larger than the lengths of the two paths $l_{\rm ref},l_d\ll l_{\rm coh}$. However, this assumption is not enough: Electrons with different energies propagate along the two paths, accumulating an energy-dependent phase difference $\theta_s=\epsilon(l_{\rm ref}-l_d)/v_F$, where $v_F$ is the Fermi velocity. As a result, the thermal averaging introduces a new lengthscale, the thermal length~\cite{doi:10.1021/nl0504585} $l_T=v_F/\pi K_BT$:
\begin{align}
&\int d\epsilon \left(-\frac{\partial f}{\partial\epsilon}\right)2\left|t_{\text{d},s}t_{\text{ref},s}\right|\cos\left[\frac{e\phi}{\hbar c}+\theta_s(\epsilon)\right]\nonumber\\
&=2\left|t_{\text{d},s}t_{\text{ref},s}\right|\cos\left[\frac{e\phi}{\hbar c}\right]\frac{l_{\rm ref}-l_d}{l_T}\frac{1}{\sinh[(l_{\rm ref}-l_d)/l_T]}.
\end{align}
Hence we also require that the difference in length between the two paths is much shorter than the thermal length~\cite{PhysRevB.80.041303} $|l_{\rm ref}-l_d|\ll l_T$. In this case, the difference in length introduces a second-order correction to the amplitude of the oscillations: $\frac{l_{\rm ref}-l_d}{l_T}\frac{1}{\sinh[(l_{\rm ref}-l_d)/l_T]}\approx1-\frac{1}{6}\left(\frac{l_{\rm ref}-l_d}{l_T}\right)^2\sim1-T^2$.

\subsection*{Open Vs. Closed Aharonov-Bohm ring}
\begin{figure}
\begin{tabular}{cc}
\includegraphics[bb=160bp 275bp 450bp 585bp,clip,scale=0.38]{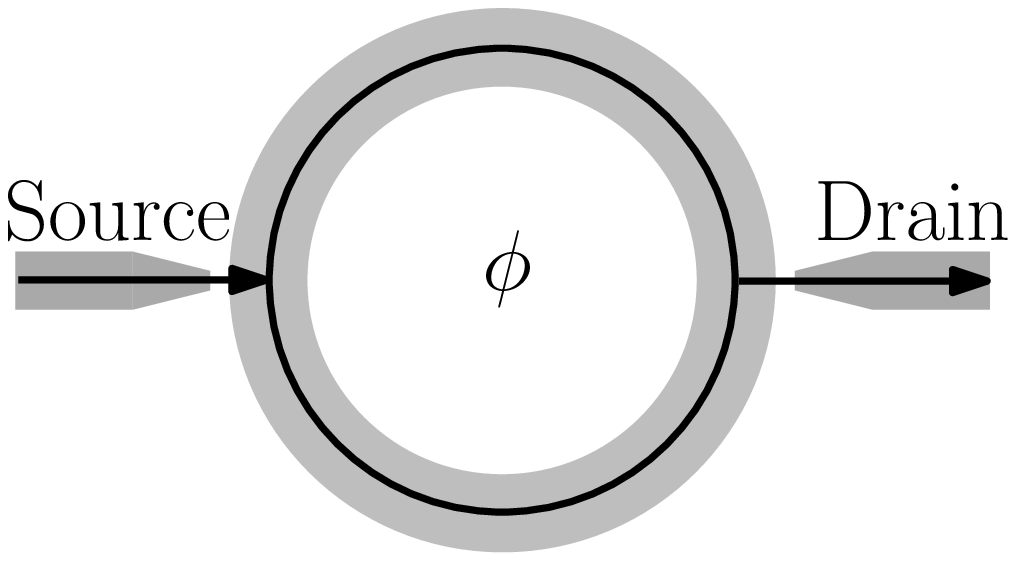} & \includegraphics[bb=137bp 275bp 488bp 585bp,clip,scale=0.38]{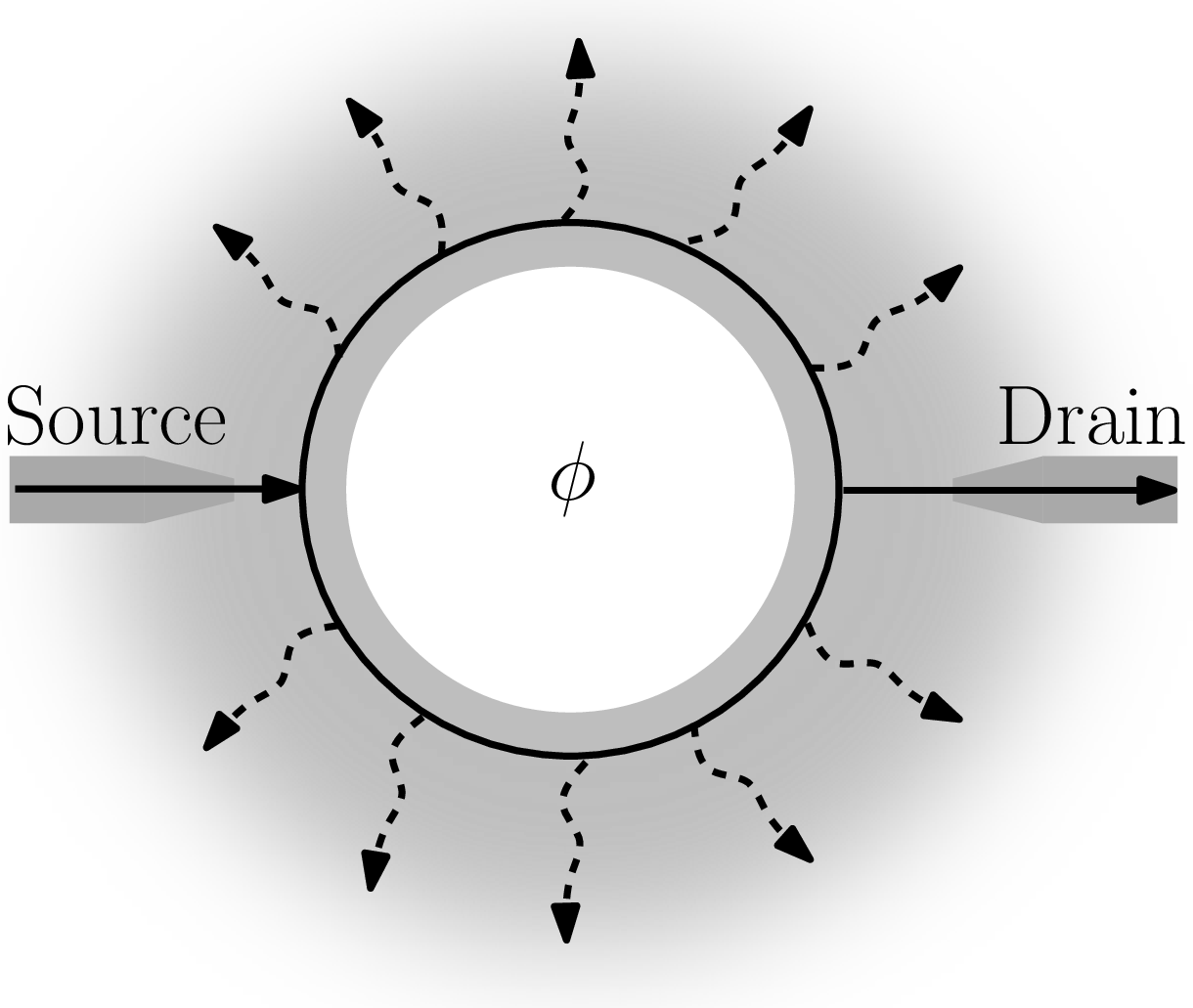} \tabularnewline (a)\mbox{\ \ Closed AB ring} & (b)\mbox{\ \ Open AB ring}
\end{tabular}
\caption{\textbf{(a)} Closed AB ring: electrons that are emitted from the source tunnel to the drain through the ring either clockwise or counter-clockwise. The two interfering paths encircle a penetrating flux, $\phi$.
Time reversal symmetry constrains the conductance:
$G(\phi)=G(-\phi).$ \textbf{(b)} Open AB ring: electrons
that propagate along the ring may leak out to side leads that are attached
to the ring. The restriction $G(\phi)=(-\phi)$ ceases
to be valid. \label{fig:1(a)}\label{fig:1(b)}}
\end{figure}

Although we will not need or discuss all its details, it is useful to have in mind a concrete physical system that realizes a two paths experiment, the AB ring. In an AB ring setup with closed geometry, as schematically drawn in Fig. \ref{fig:1(a)}(a), electrons tunnel between two leads through a conducting ring which encircles a magnetic flux. Electrons can propagate through each of the two arms of the ring, and as the two possible ways interfere, the conductance depends on the magnetic flux. Yet, there is a major difference between the closed AB ring setup and the double-slit experiment. In a naive electronic double-slit experiment picture, the phase of the interference depends continuously on the flux-tuned relative phase between the two paths.
%Cut: on $\theta$, which change continuously with the relative phase between the two paths.
In the closed AB ring, however, Onsager relations impose the restriction $G(\phi)=G(-\phi)$, which yields~\cite{Paw_1958,PhysRevLett.57.1761} $\varphi_t=\pm\pi$. This phase rigidity
has been measured~\cite{PhysRevLett.74.4047}, and although it is an interesting phenomenon by itself, it prevents a direct measurement of the phase difference between the two arms of the ring.

We can overcome this by using an open-AB-ring setup, as schematically depicted in Fig.~\ref{fig:1(b)}(b).
In such an experimental setup, that was used by Schuster \emph{et al.}~\cite{nature.385.417} and later on by others~\cite{Science.290.779,PhysRevLett.88.076601,nature.436.529,PhysRevLett.100.226601},
electrons that propagate along the ring can leak out of the ring into side leads. The loss of electrons during the propagation
through the ring relaxes the two-terminal Onsager restriction~\cite{imry2002introduction} $G(\phi)=G(-\phi)$. Although the open geometry
solves the phase rigidity problem, the intuitive double-slit
picture is not assured. In a double-slit setup, the transmissions through the two slits are independent of each other, and particles traverse the two slits only once. Therefore, we require that in the open AB ring setup, the propagation of particles along each path is independent of the details of the other path, and that there are no multiple traversals of the ring. We rely on the same features when defining the procedure for measuring $\eta$. Examples of models for open AB rings with detailed analysis of the conditions required for the realization of a double-slit setup appear in Ref.~\onlinecite{PhysRevB.66.115311} and in the appendix of Ref.~\onlinecite{PhysRevB.80.115330}.

Another difference between the AB ring and the ideal double-slit experiment is the effect that the penetrating magnetic flux has on the propagation along the two paths. In the ideal double-slit experiment, magnetic flux tunes only the relative phase of the paths.
%Cut: if a quantum dot is embedded into one of the paths, then the properties of the dot are independent of the magnetic flux.
In contrast, in a real AB ring with an embedded dot, the Kondo temperature of the dot, and the conductance through the dot, may depend on the magnetic flux. These effects of the magnetic flux on AB rings, were studied before and appear in the literature~\cite{PhysRevB.55.R7335,PhysRevLett.87.156803,PhysRevLett.88.136601,PhysRevB.72.073311,PhysRevB.72.245313,JPSJ.77.123714,PhysRevB.82.165426}.
%Cut: In this work, we use the fact that a parameter space, where these effects are weak, exists. In particular, we want
% to note that these effects are suppressed for open AB rings~\cite{PhysRevB.72.245313}. Without getting into experimental
% details of possible setups,
But these effects can be made small, particularly for open AB rings~\cite{PhysRevB.72.245313}. From now on, we thus assume an open geometry that realizes a double-slit experiment.

\section{Single-particle transmission properties and the $\mathcal{T}$-matrix}\label{sec: s-matrix discussion}
In this section, we present a more general discussion on the relation between scattering of electrons off the impurity and the conductance of the system. We relate the three measurable quantities, $G_d$, $\varphi_t$, and $\eta$, that were defined in Eq.~\eqref{eq: definition of eta and phi}, to the scattering matrix and the $\mathcal{T}$-matrix of the $\psi$-particles. In this section we mostly give the results of this discussion, whereas the full derivation appears in Appendix \ref{s-matrix appendix}.
We derive the mathematical expressions for $\varphi_t$ and $\eta$, and show that if one measures only the total conductance of the two spins together, then at $T\ll T_K$ the phase $\varphi_t$ is always equal to $\pi/2$, and it has no perturbative corrections up to order $\mathcal{O}(T/T_K)^2$ for the Fermi liquid fixed points and $\mathcal{O}(T/T_K)^{\frac{2}{2+k}}$ for the non-Fermi liquid fixed points of the $k$-channel Kondo systems.

We consider a two-path setup, and we zoom in on the path that contains the impurity. We make a distinction between the external leads (the source and the drain), and the internal leads through which the electrons propagate toward the impurity. We refer to the latter as left and right leads (see, for example, Fig.~\ref{fig:system} in Appendix \ref{two paths setup appendix}). Electrons from the source can be transmitted into the left lead, then they propagate toward the impurity. After the electrons are scattered off the impurity they can propagate along the right lead and then be transmitted out into the drain. A specific model that describes this situation is proposed and presented in Appendix \ref{two paths setup appendix}.

While the source and the drain are coupled very weakly to the internal leads (because of the losses needed to ensure each electron traverses the ring only once), the electrons in the internal leads can, in principle, interact very strongly with the impurity. Hence, in general, the left and the right leads are described by complex many-body states. A general state in the two leads can be characterized by two numbers, $n_L$ and $n_R$, measures of charge carried in each lead. There are, of course, many possible states with charges $en_L$ and $en_R$, since states with the same charges in the two leads can differ by multiple particle-hole excitations~\cite{footnote_different_channels}. We use the notation $|n_L,n_R,i\rangle$ for these states, where the index $i$ labels the possible states with charges $en_L$ and $en_R$ in the two leads.

The scattering matrix, $\mathcal{S}$, connects incoming and outgoing states in the leads
\begin{equation}
|n'_L,n'_R,j\rangle_{\rm out}=\mathcal{S}^{n'_L,n'_R,j}_{n_L,n_R,i}|n_L,n_R,i\rangle_{\rm in}\;.\label{eq: s-matrix-definition_bulk}
\end{equation}
Charge conservation imposes $n'_L+n'_R=n_L+n_R=m$, so $\mathcal{S}$ is a block-diagonal matrix, as sectors with different integer value $m$, are not mixed. Since the source and the drain are coupled very weakly to the internal leads, in the limit of zero source-drain bias voltage at low temperature, we assume that only one particle at a time is launched from the external leads toward the impurity. Hence, we focus only on the block $m=1$ of the $\mathcal{S}$-matrix.
When a single electron is sent from the source, through the left lead, into the impurity, there are three possible options:
\begin{itemize}
  \item The electron is reflected back to the left lead,
  \item The electron is transmitted to the right lead,
  \item A complex many-body state is produced, where a total charge $ne$ is transmitted to the right lead and a charge $(1-n)e$ is reflected to the left lead ($n=0,\pm1,\pm2...$) .
\end{itemize}

We want to distinguish between the elastic single-particle scattering processes and the scattering processes that involve many-body states. We therefore use the following notation:
%we denote by $|L\rangle_{\rm in}$ and $|L\rangle_{\rm out}$ the incoming and outgoing single-electron states in the left %lead, and similarly $|R\rangle_{\rm in}$ and $|R\rangle_{\rm out}$ in the right lead.
we denote by $|L\rangle$ the incoming or outgoing single-electron states in the left lead, and similarly $|R\rangle$ in the right lead.
In the notation $|n_L,n_R,i\rangle$,
\begin{equation}
|L\rangle=|1,0,0\rangle\;, \;\;\;\;\;\;\;\;\;\;|R\rangle=|0,1,0\rangle\;,\label{eq: definition L,R}
\end{equation}
where we arbitrarily choose $i=0$ for the single-particle states with total charge one.
The many-body states (also with total charge one) are denoted by $|\chi^i_n\rangle$, where
\begin{equation}
|\chi^i_n\rangle=|1-n,n,i\rangle\;.
\end{equation}
We use the following notation for the $\mathcal{S}$-matrix elements that connect incoming single-particle states with outgoing single-particle states:
\begin{equation}
\mathcal{S}^{1,0,0}_{1,0,0}=r\;,\;\;\mathcal{S}^{1,0,0}_{0,1,0}=t\;,\;\;\mathcal{S}^{0,1,0}_{0,1,0}=r'\;,\;\;\mathcal{S}^{0,1,0}_{1,0,0}=t'\;.
\end{equation}
The matrix elements that connect single-particle states with many-body states are:
\begin{align}
&\mathcal{S}^{1-n,n,i}_{1,0,0}=B_L^{ni}\;,\;\;\;\;\;\;\;\mathcal{S}^{1-n,n,i}_{0,1,0}=B_R^{ni}\;,\\
&\mathcal{S}^{1,0,0}_{1-n,n,i}=\left(A_L^{ni}\right)^*\;,\;\;\mathcal{S}^{0,1,0}_{1-n,n,i}=\left(A_R^{ni}\right)^*\;.\label{eq: definition A}
\end{align}
Schematically, the $n_L+n_R=1$ block of the $\mathcal{S}$-matrix is
\begin{equation}
\left(
  \begin{array}{c}
    |L\rangle_{\rm out} \\
    |R\rangle_{\rm out} \\
    |\boldsymbol{\chi}\rangle_{\rm out} \\
  \end{array}
\right)=\left(
  \begin{array}{cccccccc}
    r & t' &  \boldsymbol{A_L}^{\dagger}\\
    t & r' &  \boldsymbol{A_R}^{\dagger}\\
    \boldsymbol{B_L} & \boldsymbol{B_R} & \boldsymbol{C}\\
  \end{array}
\right)
\left(
  \begin{array}{c}
    |L\rangle_{\rm in} \\
    |R\rangle_{\rm in} \\
    |\boldsymbol{\chi}\rangle_{\rm in} \\
  \end{array}
\right)\;,
\end{equation}
where the matrix $\boldsymbol{C}$ denotes the matrix elements of $\mathcal{S}$ that connect incoming many-body states with outgoing many-body states. Here we don't include spin, but generalization of what follows to spinful electrons is straightforward.

Consider now the average current at the right lead. The current is carried either by transmitted charge (from the left), or by reflected charge
\begin{align}
I&=\frac{e}{h}\int d\epsilon\left[f_l(\epsilon)\left(|t|^2+\sum_{n,i} n\left|B_L^{ni}\right|^2\right)\right.\nonumber\\
&\left.+f_r(\epsilon)\left(|r'|^2+\sum_{n,i} n\left|B_R^{ni}\right|^2-1\right)\right]\; .\nonumber
\end{align}
Using the unitarity of the large many-body $\mathcal{S}$-matrix we can write the conductance through the impurity as
\begin{equation}
G=\frac{e^2}{h}\int d\epsilon\left(-\frac{\partial f}{\partial\epsilon}\right)\left(|t|^2+\sum_{n,i} n\left|B_L^{ni}\right|^2\right).\label{eq: Conductance_section}
\end{equation}
The coherent part of the conductance is obtained directly from Eq.~\eqref{eq: Conductance_section}
\begin{equation}
G_{\rm coh}=\frac{e^2}{h}\int d\epsilon\left(-\frac{\partial f}{\partial\epsilon}\right)|t|^2\; .\label{eq: coherent conductance}
\end{equation}
The contribution of the incoherent processes, where the single particles are scattered into many-body states, is
\begin{equation}
G_{\rm incoh}=\frac{e^2}{h}\int d\epsilon\left(-\frac{\partial f}{\partial\epsilon}\right)\sum_{n,i} n\left|B_L^{ni}\right|^2\; .\label{eq: incoherent conductance}
\end{equation}

Suppose now, that there is a unitary transformation that mixes the two leads and  block-diagonals the $n_L+n_R=1$ block of the $\mathcal{S}$-matrix. Physically, it means that there is a linear combination of the two leads, $\xi=-\sin(\alpha)L+\cos(\alpha)R$ , which is decoupled both from the impurity and from the orthogonal combination of the leads, $\psi=\cos(\alpha)L+\sin(\alpha)R$. This is the case, for example, in the Anderson model for a single-level quantum dot that is coupled to two leads.
This simplification breaks down in many-level quantum dots~\cite{Glazman2005427}, so in this paper we assume for simplicity a single-level quantum dot.

The single-$\psi$-particle matrix element of the $\mathcal{S}$-matrix in the new basis is
\begin{equation}
\mathcal{S}_{\psi\psi}=1+\frac{t}{\cos(\alpha)\sin(\alpha)}\;.\nonumber
\end{equation}
Moreover, the fact that $\xi$ is a free decoupled field imposes the following relation
\begin{equation}
\sum_{n,i}n\left|B_L^{ni}\right|^2=\cos^2(\alpha)\sin^2(\alpha)(1-|\mathcal{S}_{\psi\psi}|^2)\;.
\end{equation}
Using the definition $\mathcal{S}=1+i\mathcal{T}$ for the $\mathcal{T}$-matrix, we get the known result~\cite{PhysRevB.64.045328,springerlink:10.1007/978-3-540-72632-6_2,review_DO} for the conductance through the impurity
\begin{equation}
G_{\rm d}=\frac{e^2}{h}\frac{\sin^2(2\alpha)}{4}\int d\epsilon\left(-\frac{\partial f}{\partial\epsilon}\right)2Im\left\{\mathcal{T}_{\psi\psi}\right\}\;.
\end{equation}
The ratio of the coherent part to the total conductance is
\begin{equation}
G_{\rm coh}/G_{\rm d}=\frac{\int d\epsilon\left(-\frac{\partial f}{\partial\epsilon}\right)\left|\mathcal{T}_{\psi\psi}\right|^2}{\int d\epsilon\left(-\frac{\partial f}{\partial\epsilon}\right)2Im\left\{\mathcal{T}_{\psi\psi}\right\}}\;.
\end{equation}
%Notice that $\mathcal{T}_{\psi\psi}$ is related to the transmission $t$ via
%\begin{equation}
%\mathcal{T}_{\psi\psi}=-i\frac{t}{\cos(\alpha)\sin(\alpha)}\;.
%\end{equation}

\subsection{Normalized visibility}
There is no way to measure directly the contribution of the single-particle processes to the conductance. Namely, there is no direct measurement of $G_{\rm coh}/G_{\rm d}$ . However, a two-path experiment gives access to the transmission amplitude, $t$. If in addition to the impurity, the two leads are connected via an independent free reference arm, then the flux-dependent part of the conductance is $G_{\rm flux}=\int d\epsilon\left(-\frac{\partial f}{\partial\epsilon}\right)2Re\left\{t_{\rm ref}te^{i\frac{e\phi}{\hbar c}}\right\}$ . Since $|t_{\rm ref}|$ can be extracted from the conductance of the reference arm when the other arm closed off, $t$ is accessible from the flux-dependent conductance.

While $G_{\rm coh}$ is proportional to the thermally-averaged value of the transmission squared [see Eq.~\eqref{eq: coherent conductance}], $G_{\rm flux}$ is proportional to the thermally-averaged value of the transmission, $\int d\epsilon\left(-\frac{\partial f}{\partial\epsilon}\right)t(\epsilon)$. The normalized visibility that we have defined in Eq.~(\ref{eq: definition of eta and phi}) is therefore slightly different from $G_{\rm coh}/G_{\rm d}$
\begin{equation}
\eta=\frac{\left|\int d\epsilon\left(-\frac{\partial f}{\partial\epsilon}\right)\mathcal{T}_{\psi\psi}\right|^2}{\int d\epsilon\left(-\frac{\partial f}{\partial\epsilon}\right)2Im\left\{\mathcal{T}_{\psi\psi}\right\}}\;.\label{eq: eta_no_spin}
\end{equation}
Although $G_{\rm coh}/G_{\rm d}$ is closely related to the measurable quantity $\eta$, they are identical only at zero temperature, or where $\mathcal{T}_{\psi\psi}$ is independent of the energy.

\subsection{Transmission phase}
The phases of $t$ and $\mathcal{T}_{\psi\psi}$ are related to the phase shift of the scattering theory of the $\psi$-particles. If we write $\mathcal{S}_{\psi\psi}=|\mathcal{S}_{\psi\psi}|e^{2i\delta_{\psi}}$ , then
\begin{equation}
\arg(\mathcal{T}_{\psi\psi})=\arctan\left(\frac{1-|\mathcal{S}_{\psi\psi}|\cos(2\delta_{\psi})}{|\mathcal{S}_{\psi\psi}|\sin(2\delta_{\psi})}\right).\label{eq: phase shift vs transmission phase}
\end{equation}
The phase $\arg(\mathcal{T}_{\psi\psi})$ yields the value $\delta_{\psi}$ for $|\mathcal{S}_{\psi\psi}|\rightarrow1$ and $\pi/2$ in the limit $|\mathcal{S}_{\psi\psi}|\rightarrow0$.
The transmission phase is the phase of the thermally averaged $\mathcal{T}$-matrix
\begin{equation}
\varphi_t=\arg\left\{\int d\epsilon\left(-\frac{\partial f}{\partial\epsilon}\right)\mathcal{T}_{\psi\psi}(\epsilon)\right\}.\label{eq: varphi_t_no_spin}
\end{equation}

\subsection{The $\pi/2$ phase-lock of the transmission through Kondo impurities at $T\ll T_K$}
The flux-dependent part of the conductance, $G_{\rm flux}$, depends on the average value of $\mathcal{T}_{\psi\psi}$. Until now, the averaging was over different incoming energies (thermal averaging). When we add the spin degree of freedom, we average $\mathcal{T}_{\psi\psi}$ also over spin. This is because in $G_{\rm flux}$, we sum over the two spins
\begin{align}
G_{\rm flux}&=-\sum_s\int d\epsilon\frac{\partial f}{\partial\epsilon}2Re\left\{t_{\rm ref}t_se^{i\frac{e\phi}{\hbar c}}\right\}\\
=&-\sum_s\int d\epsilon\frac{\partial f}{\partial\epsilon}2Re\left\{i\cos(\alpha)\sin(\alpha)t_{\rm ref}\mathcal{T}_{s,\psi\psi}e^{i\frac{e\phi}{\hbar c}}\right\}\; .\nonumber
\end{align}
We have assumed that $t_{\rm ref}$ is independent of the spin.
If the system is spin-symmetric, $\mathcal{T}_{\uparrow,\psi\psi}=\mathcal{T}_{\downarrow,\psi\psi}\equiv\mathcal{T}_{\psi\psi}$ can be extracted from $G_{\rm flux}$.
The normalized visibility in this case is
\begin{align}
\eta=\frac{\left|\int d\epsilon\left(-\frac{\partial f}{\partial\epsilon}\right)\mathcal{T}_{\psi\psi}\right|^2}{\int d\epsilon\left(-\frac{\partial f}{\partial\epsilon}\right)2Im\left\{\mathcal{T}_{\psi\psi}\right\}}\;,
\end{align}
and the transmission phase is
\begin{equation}
\varphi_t=\arg\left\{\int d\epsilon\left(-\frac{\partial f}{\partial\epsilon}\right)\mathcal{T}_{\psi\psi}(\epsilon)\right\}\;.
\end{equation}

In the absence of spin-symmetry, $G_{\rm flux}$ does not necessarily give us access to $\mathcal{T}_{s,\psi\psi}$. To see this, consider the simple case where all the particles are scattered into single particles, namely, $|\mathcal{S}_{\psi\psi}|=1$ for both spins. This situation describes, for example, the Fermi-liquid fixed points of 1CK or 2CK with an applied magnetic field. In this case, $\mathcal{T}_{s,\psi\psi}=i(1-e^{2i\delta_{\psi s}})=2\sin(\delta_{\psi s})e^{i\delta_{\psi s}}$. In the Kondo case, the system has the following particle-hole symmetry [see, for example, the Hamiltonian in Eq.\eqref{eq: SCK H}]
\begin{equation}
\psi_{ks}\rightarrow\psi^{\dagger}_{-k,-s}\;,\label{eq: p-h symmetry}
\end{equation}
that enforces~\cite{J.low.temp.phys.17.31,0953-8984-16-16-R01} $\delta_{\psi\uparrow}(\epsilon)=-\delta_{\psi\downarrow}(-\epsilon)$~. The transmission phase at zero temperature is
\begin{equation}
\varphi_t=\arg\left[\sin(\delta_{\psi\uparrow})(e^{i\delta_{\psi\uparrow}}-e^{-i\delta_{\psi\uparrow}})\right]=\frac{\pi}{2}\;,\label{eq: phase lock pi/2}
\end{equation}
and the normalized visibility at zero temperature
\begin{equation}
\eta=\frac{\left|\sin(\delta_{\psi\uparrow})(e^{i\delta_{\psi\uparrow}}-e^{-i\delta_{\psi\uparrow}})\right|^2}{2Im\left\{\sin(\delta_{\psi\uparrow})(e^{i\delta_{\psi\uparrow}}-e^{-i\delta_{\psi\uparrow}})\right\}}=\sin^2(\delta_{\psi\uparrow})\;.\label{eq: normalized visibility two spins}
\end{equation}
We see that the transmission phase is locked at $\pi/2$~, independent of the phases of $\mathcal{T}_{s,\psi\psi}$~. We also see that the normalized visibility can be smaller than one, even though all the scattering processes are single-particle to single-particle scattering. Interestingly, information about the phases of $\mathcal{T}_{s, \psi\psi}$ (the phase shifts of the scattering theory), is now encoded in $\eta$~.

There are two ways to extract $\mathcal{T}_{\psi\psi}$ despite the $\pi/2$ phase-lock of Kondo impurities: either to use the normalized visibility to extract the phase shift, or to measure the transmission of each spin separately. A concrete way to realize Kondo impurities in quantum dots, with access to each spin separately was proposed by some of the present authors in Ref.~\onlinecite{PhysRevLett.106.106401}.

Note that if the Kondo impurity is realized with a quantum dot, the particle-hole symmetry~\eqref{eq: p-h symmetry} is exact only if the dot is tuned by the gate voltage to the middle of the Coulomb valley~\cite{PhysRevLett.87.216601,PhysRevB.69.115316}. Weakly breaking the particle-hole symmetry adds a spin-independent contribution to the phase shift, $\delta_{\psi s}\rightarrow\delta_0+\delta_{\psi s}$ (this is true both for Fermi liquid cases and the non-Fermi liquid case of the 2CK~\cite{PhysRevB.69.115316}). For $\delta_0\ll\delta_{\psi s}$ it leads to small corrections of Eqs.~\eqref{eq: phase lock pi/2} and \eqref{eq: normalized visibility two spins}:
\begin{align}
&\varphi_t=\frac{\pi}{2}+[1-\cot^2(\delta_{\psi\uparrow})]\delta_0+\mathcal{O}\left(\frac{\delta_0}{\delta_{\psi\uparrow}}\right)^3\;,\\
&\eta=\sin^2(\delta_{\psi\uparrow})+\cos(2\delta_{\psi\uparrow})\cot^2(\delta_{\psi\uparrow})\delta_0^2+\mathcal{O}\left(\frac{\delta_0}{\delta_{\psi\uparrow}}\right)^4\;.\nonumber
\end{align}

%
%%%%%%%%%%%%%%%%%%%%%%%%%%%%%%%%%%%%%%%%%%%%%%%%%%%%%%%%%%%%%%%%%%%%%%%%%%%%%%%%%%%%%%%%%%%%%%%%%%%%%%%%%%%%%%%%%%%%%%
%%%%%%%%%%%%%%%%%%%%%%%%%%%%%%%%%%%%%%%%%%%%%%%%%%%%%%%%%%%%%%%%%%%%%%%%%%%%%%%%%%%%%%%%%%%%%%%%%%%%%%%%%%%%%%%%%%%%%%
%%%%%%%%%%%%%%%%%%%%%%%%%%%%%%%%%%%%%%%%%%%%%%%%%%%%%%%%%%%%%%%%%%%%%%%%%%%%%%%%%%%%%%%%%%%%%%%%%%%%%%%%%%%%%%%%%%%%%%
%%%%%%%%%%%%%%%%%%%%%%%%%%%%%%%%%%%%%%%%%%%%%%%%%%%%%%%%%%%%%%%%%%%%%%%%%%%%%%%%%%%%%%%%%%%%%%%%%%%%%%%%%%%%%%%%%%%%%%
%%%%%%%%%%%%%%%%%%%%%%%%%%%%%%%%%%%%%%%%%%%%%%%%%%%%%%%%%%%%%%%%%%%%%%%%%%%%%%%%%%%%%%%%%%%%%%%%%%%%%%%%%%%%%%%%%%%%%%
%%%%%%%%%%%%%%%%%%%%%%%%%%%%%%%%%%%%%%%%%%%%%%%%%%%%%%%%%%%%%%%%%%%%%%%%%%%%%%%%%%%%%%%%%%%%%%%%%%%%%%%%%%%%%%%%%%%%%%
%%%%%%%%%%%%%%%%%%%%%%%%%%%%%%%%%%%%%%%%%%%%%%%%%%%%%%%%%%%%%%%%%%%%%%%%%%%%%%%%%%%%%%%%%%%%%%%%%%%%%%%%%%%%%%%%%%%%%%
%%%%%%%%%%%%%%%%%%%%%%%%%%%%%%%%%%%%%%%%%%%%%%%%%%%%%%%%%%%%%%%%%%%%%%%%%%%%%%%%%%%%%%%%%%%%%%%%%%%%%%%%%%%%%%%%%%%%%%
%%%%%%%%%%%%%%%%%%%%%%%%%%%%%%%%%%%%%%%%%%%%%%%%%%%%%%%%%%%%%%%%%%%%%%%%%%%%%%%%%%%%%%%%%%%%%%%%%%%%%%%%%%%%%%%%%%%%%%
%%%%%%%%%%%%%%%%%%%%%%%%%%%%%%%%%%%%%%%%%%%%%%%%%%%%%%%%%%%%%%%%%%%%%%%%%%%%%%%%%%%%%%%%%%%%%%%%%%%%%%%%%%%%%%%%%%%%%%
%
\section{Results}\label{sec:results}
In this section, we present the results of the transmission phase $\varphi_t$, and normalized visibility $\eta$ of Kondo impurities [both were defined in Eq. (\ref{eq: definition of eta and phi})]. We focus on the 1CK impurity and the 2CK impurity, since there are concrete realizations of such impurities with quantum dots, and only quote the results for the general k-channel Kondo. In the 2CK case, we consider both its non-Fermi liquid fixed point, and its Fermi liquid fixed points, reached by turning on a finite magnetic field or a finite channel anisotropy.

\subsection{Single channel Kondo}
In the 1CK case, the $\mathcal{T}_{s, \psi\psi}$ matrix element, up to second order in $1/T_K$, is~\cite{PhysRevB.48.7297}
\begin{equation}
\mathcal{T}_{s,\psi\psi}\left(\epsilon\right)=i\left[2+i\frac{2\epsilon}{T_K}-3\left(\frac{\epsilon}{T_K}\right)^{2}-\left(\frac{\pi T}{T_K}\right)^{2}\right]\;.
\end{equation}

Since $\int\epsilon d\epsilon\left(-\frac{\partial f}{\partial\epsilon}\right)=0$, then $\int d\epsilon\left(-\frac{\partial f}{\partial\epsilon}\right)\mathcal{T}_{s,\psi\psi}(\epsilon)$ is purely imaginary, therefore the transmission phase is
\begin{equation}
\varphi_t=\frac{\pi}{2}+\mathcal{O}\left(T/T_K\right)^3\;.\label{eq:1CK phase shift}
\end{equation}
The transmission phase matches the scattering phase shift of the 1CK (up to $T/T_K$ corrections) when potential scattering is neglected.
The normalized visibility
\begin{equation}
\eta=1-\left(\frac{\pi T}{T_K}\right)^2+\mathcal{O}\left(T/T_K\right)^3\; .
\end{equation}
Two mechanisms reduce the nonzero-temperature normalized visibility, elastic scattering with energy-dependent phase shift, $\delta_\psi(\epsilon)=\pi/2+\epsilon/T_K$, and the appearance of inelastic scattering. Both are allowed by the dominant irrelevant operator near the 1CK fixed point~\cite{PhysRevB.48.7297}.

\subsubsection{Finite magnetic field}
At zero magnetic field, the $\mathcal{T}$-matrix is independent of spin (\emph{i.e.}, $\mathcal{T}_{\uparrow,\psi\psi}=\mathcal{T}_{\downarrow,\psi\psi}$), because of the symmetry between the two spins. Therefore, the transmission phase and the normalized visibility of the spin-summed conductance, are the same as the transmission phase and the normalized visibility of each spin separately. However, when a magnetic field is applied, the $\mathcal{T}$-matrix becomes spin-dependent. Hence, the transmission phase and the normalized visibility of each spin are, in general, different from each other and from the measured (spin-summed) quantities.

Consider, for example, the zero temperature case, where, as long as $B\ll T_K$, the system is described by a Fermi liquid theory, so~\cite{0953-8984-16-16-R01} $\mathcal{T}_{s,\psi\psi}=i(1-e^{2i\delta_{\psi s}})$. As we discussed in section \ref{sec: s-matrix discussion}, the particle hole symmetry $\psi_{ks}\rightarrow\psi^{\dagger}_{-k,-s}$, enforces $\delta_{\psi \uparrow}(\epsilon)=-\delta_{\psi \downarrow}(-\epsilon)$. In this case,
\begin{equation}
\delta_{\psi s}(0)=\left(\frac{\pi}{2}-\alpha_s\frac{B}{T_K}\right)\;,
\end{equation}
where $\alpha_\uparrow=1$, and $\alpha_\downarrow=-1$.
Notice that since $\delta_{\psi s}$ is \emph{half} of the phase of $\mathcal{S}_{\psi\psi}$, it is defined up to $\pm\pi$.
As we measure the conductance of the two spins together, the total transmission phase $\varphi_t=\pi/2$ independent of $B$ [see Eq.~\eqref{eq: phase lock pi/2}], and the normalized visibility is less than one, $\eta=\sin^2(\frac{\pi}{2}-\frac{B}{T_K})\approx1-(\frac{B}{T_K})^2$ [see Eq.~\eqref{eq: normalized visibility two spins}], even though all the scattering processes are single-particle to single-particle scattering.

A possible way to overcome this $\pi/2$ phase-lock of the transmission phase, is to measure the conductance of a distinct spin~\cite{PhysRevLett.106.106401}. The distinct spin transmission phase at zero temperature would simply be $\delta_{\psi s}$, and there is a $\frac{2B}{T_K}$ difference between the spin up and spin down phases. The normalized visibility of each distinct spin would be $\eta=1$, as we expect for a Fermi liquid fixed point.

\subsection{Two channel Kondo}

In the 2CK case, two disconnected channels interact with the impurity. We consider a case where we can measure the transport in one of the channels, and there is no charge transfer between the different channels (this was the case, for example, in the experimental setup of Ref.~\onlinecite{nature.446.167}). Notice that in this case, the index $i$ in the states $|n_L,n_R,i\rangle$ [see, for example, equation~\eqref{eq: s-matrix-definition_bulk}], labels states with different particle-hole excitations in the leads and also states with different excitations in the other channel.

If the two channels are equally coupled to the impurity, then the system flows to a non-Fermi liquid fixed point. In this case, up to first order in $1/\sqrt{T_K}$,
the matrix element $\mathcal{T}_{s, \psi\psi}$ is~\cite{PhysRevB.48.7297}
\begin{equation}
\mathcal{T}_{s,\psi\psi}\left(\epsilon\right)=i\left(1-3\lambda\sqrt{\pi T}I(\epsilon)\right),
\end{equation}
where
\begin{equation}
I(\epsilon)=\int_{0}^{1}du\left(u^{-\frac{i\epsilon}{2\pi T}}F_{21}(u)\sqrt{\frac{1-u}{u}}-\frac{4}{\pi}\frac{1}{\sqrt{u}(1-u)^{\frac{3}{2}}}\right)\;.\label{eq: definition of I(e)}
\end{equation}
$\lambda\sim1/\sqrt{T_K}$ is the strength of the leading irrelevant operator near the 2CK fixed point, and $F_{21}(u)$ is the hypergeometric function $F_{21}(u)\equiv\frac{1}{2\pi}\int_{o}^{2\pi}\frac{d\theta}{\left(u+1-2\sqrt{u}\cos\theta\right)^{\frac{3}{2}}}$~.

The thermally averaged value of $\mathcal{T}_{s,\psi\psi}$ is
\begin{equation}
\int d\epsilon\left(-\frac{\partial f}{\partial\epsilon}\right)\mathcal{T}_{s,\psi\psi}(\epsilon)=i\left(1+4\lambda\sqrt{\pi T}\right).
\end{equation}
Since $\int d\epsilon\left(-\frac{\partial f}{\partial\epsilon}\right)\mathcal{T}_{s,\psi\psi}(\epsilon)$ is purely imaginary, the transmission phase
\begin{equation}
\varphi_t=\frac{\pi}{2}+\mathcal{O}\left(T/T_K\right)\;.\label{eq:2CK phase shift}
\end{equation}
The normalized visibility is
\begin{equation}
\eta=\frac{1}{2}\left(1+4\lambda\sqrt{\pi T}\right)+\mathcal{O}\left(T/T_K\right)\;.\label{eq:2CK normalized visibility}
\end{equation}
These results are not surprising, since at zero temperature, there are no single $\psi$-particle to single $\psi$-particle scattering processes at the non-Fermi liquid fixed point. Thus, $\mathcal{S}_{\psi\psi}=0$ for both spins, and hence $\varphi_t=\pi/2$ [see Eq.~\eqref{eq: phase shift vs transmission phase}]. Since in this case $\mathcal{T}_{\psi\psi}=i$, we find a normalized visibility $\eta=1/2$, indicating that half of the conductance is carried by elastic single-particle scattering~\cite{PhysRevLett.93.107204,PhysRevB.75.235112}.

The sign of $\lambda$ depends on the initial strength of the Kondo coupling. $\lambda$ is positive for strong coupling, and negative for weak coupling~\cite{PhysRevB.48.7297}.
The normalized visibility can, in principle, be \emph{enhanced} by nonzero temperature, unlike the usual case where the temperature reduces interference effects. The enhancement of the normalized visibility is due to the fact that the nonzero temperature allows single $\psi$-particles scattering off the impurity ($s_{\psi\psi}\neq0$).

\subsubsection{Finite magnetic field and finite channel anisotropy}
The non-Fermi liquid fixed point is unstable, since finite magnetic field and finite channel anisotropy turn on relevant perturbations near the non-Fermi liquid fixed point~\cite{PhysRevB.45.7918}.
In the presence of such perturbations, the system flows under renormalization group to a Fermi liquid fixed point, at zero temperature, rather than the non-Fermi liquid one.
In the case of channel anisotropy, the channel which is coupled more strongly to the dot flows to the 1CK-like fixed point, and the other channel flows to a free-electrons-like fixed point.
Under a finite magnetic field, the system flows to a Fermi liquid fixed point which is different from the 1CK fixed point.

In this subsection we study the 2CK case under these two possible perturbations.
At zero temperature, $\mathcal{T}_{s,\psi\psi}$ is given by~\cite{PhysRevLett.106.147202,arXiv:1203.4456v1}
\begin{equation}
\mathcal{T}_{s,\psi\psi}(\epsilon)=i\left[1-\frac{-(\nu\Delta\sqrt{T_K})+i\alpha_s(\frac{c_BB_z}{\sqrt{T_K}})}{\sqrt{(\nu\Delta\sqrt{T_K})^2+(\frac{c_BB_z}{\sqrt{T_K}})^2}}\mathcal{G}(\epsilon/T^*)\right],
\end{equation}
where $\Delta$ is the difference between the coupling strengths of the two channels, and $c_B$ is a dimensionless number of order one. $T^*\sim T_K(\nu\Delta)^2+(c_BB)^2/T_K$ is an energy scale that characterizes the flow away from the non-Fermi liquid fixed point. $\mathcal{G}(x)=\frac{2}{\pi}K(ix)$, where $K(x)$ is the complete elliptic integral of the first kind. $\alpha_\uparrow=1$, $\alpha_\downarrow=-1$, and we have assumed $\vec{B}=B_z$.
At zero temperature, the averaged value of $\mathcal{T}_{\psi\psi}$ is
\begin{equation}
\frac{1}{2}\sum_s \mathcal{T}_{s,\psi\psi}=i\left[1-\frac{-(\nu\Delta\sqrt{T_K})}{\sqrt{(\nu\Delta\sqrt{T_K})^2+(\frac{c_BB_z}{\sqrt{T_K}})^2}}\right]\;.
\end{equation}
Thus, for $\Delta=0$, $\langle\mathcal{T}_{\psi\psi}\rangle=i$. Hence, the transmission phase is $\varphi_t=\pi/2$ and the normalized visibility is $\eta=1/2$ even for $B\neq0$, where all the electrons are elastically scattered with a phase $\delta_{\psi,s}=\alpha_s\pi/4$. A spin-resolved measurement, however, would lead to $\varphi_t=\alpha_s\pi/4$ and $\eta=1$, since for $\Delta=0$
\begin{equation}
\int d\epsilon\left(-\frac{\partial f}{\partial\epsilon}\right)\mathcal{T}_{s,\psi\psi}=i\left(1-i\alpha_s\right)\;.
\end{equation}

In Table \ref{table}, we summarize the results for the zero temperature normalized visibility and transmission phase for the various relevant perturbations, where we define
\begin{eqnarray}
&&\cos(\gamma)\equiv\frac{\nu|\Delta|\sqrt{T_K}}{\sqrt{(\nu\Delta\sqrt{T_K})^2+(\frac{c_BB_z}{\sqrt{T_K}})^2}}\;,\\
&&\sin(\gamma)\equiv\frac{c_BB_z/\sqrt{T_K}}{\sqrt{(\nu\Delta\sqrt{T_K})^2+(\frac{c_BB_z}{\sqrt{T_K}})^2}}\;.
\end{eqnarray}

\emph{Channel anisotropy}. Recall that we are measuring the conductance through one of the channels. At zero magnetic field, if $\Delta>0$, the $\psi$-particles form together with the impurity a singlet, while the electrons in the other channel are simply free. Thus, $\eta$ and $\varphi_t$ are the same as in the 1CK case. On the other hand, if $\Delta<0$, the electrons in the other channel form a singlet with the impurity, and the $\psi$-particles are free. Therefore at zero temperature the conductance through the impurity, the dot, is zero. In this case there is no interference, and hence, $\eta=0$ and $\varphi_t$ is not defined. Although this is a Fermi liquid, $\eta<1$ near this fixed point since most of the charge is reflected. To explain it we now discuss the nonzero-temperature case.

At nonzero temperature, the $\Delta<0$ case should be treated more delicately. Up to second order in $1/T^*$, $\mathcal{T}_{s,\psi\psi}$ is~\cite{arXiv:1203.4456v1}
\begin{equation}
\mathcal{T}_{s,\psi\psi}(\epsilon)=\frac{\epsilon}{4T^*}+i\frac{9}{64}\left(\frac{\epsilon}{T^*}\right)^2+i\frac{7}{64}\left(\frac{\pi T}{T^*}\right)^2\;.
\end{equation}
Most of the charge is reflected and only a small amount of charge can be transmitted, either elastically or inelastically. This is similar to the 1CK case, where at nonzero temperature most of the charge is transmitted, and only a small part is reflected either elastically or inelastically.
Up to second order in $1/T^*$, the portion of elastic transmission through the impurity out of all scattering events of incoming particles with energy $\epsilon$ is
\begin{equation}
\frac{\left|\mathcal{T}_{s,\psi\psi}(\epsilon)\right|^2}{2Im\left\{\mathcal{T}_{s,\psi\psi}(\epsilon)\right\}}=\frac{2/9}{1+\frac{7}{9}\left(\frac{\pi T}{\epsilon}\right)^2}\;.
\end{equation}
In the $\epsilon\gg T$ limit, $2/9$ of the charge is transmitted elastically. The phase that the particles accumulate in this limit is proportional to $\epsilon$, $\varphi_t(\epsilon)\approx\frac{9\epsilon}{16T^*}$.
The thermal averaging, however, has a crucial effect in this limit. The thermally-averaged $\mathcal{T}$-matrix ,$\langle\mathcal{T}_{\psi\psi}\rangle=i\frac{5}{32}\left(\frac{\pi T}{T^*}\right)^2$, is purely imaginary and proportional to $T^2$, and therefore
\begin{equation}
\eta(T)=5\left(\frac{\pi T}{8T^*}\right)^2\;,\;\;\;\;\; \varphi_t=\pi/2\;.
\end{equation}

\emph{Finite magnetic field}. At finite magnetic field, we see that in order to access the phase shift of the $\psi$-particles, $\delta_{\psi s}$, one needs to measure each spin separately. Notice that at $\Delta\rightarrow0$ ($\gamma\rightarrow\pi/2$), the spin-averaged normalized visibility and the transmission phase are the same as in the non-Fermi liquid fixed point ($B=0, \Delta=0$): $\eta=1/2$ and $\varphi_t=\pi/2$. In order to distinguish the Fermi-liquid fixed points from the non-Fermi liquid fixed point, one can measure the temperature dependence of the conductance through the impurity. Non trivial $\sqrt{T}$-dependence indicates a non-Fermi liquid fixed point. Alternatively, as we already mentioned, spin dependent measurements of $\eta_s$ and $\varphi_{ts}$ give different results for the Fermi liquid and the non-Fermi liquid fixed points.
\begin{center}
\begin{table}
  \caption{Zero temperature normalized visibility and transmission phase for various relevant perturbations.}
\begin{tabular}{|c|c|c|c|c|}
  \hline
    & $\eta_s$ & $\varphi_{ts}$ & $\eta$ & $\varphi_{t}$ \\\hline
  $B=0,\Delta=0$ & 1/2 & $\pi/2$ & 1/2 & $\pi/2$ \\
  $B=0,\Delta>0$ & 1 & $\pi/2$ & 1 & $\pi/2$ \\
  $B=0,\Delta<0$ & 0 & - & 0 & - \\
  $B\neq0,\Delta=0$ & 1 & $\alpha_s\pi/4$ & $1/2$ & $\pi/2$ \\
  $B\neq0,\Delta>0$ & 1 & $\alpha_s(\pi/2-\gamma/2)$ & $\cos^2(\gamma/2)$ & $\pi/2$ \\
  $B\neq0,\Delta<0$ & 1 & $\alpha_s\gamma/2$ & $\sin^2(\gamma/2)$ & $\pi/2$ \\
  \hline
\end{tabular}
\label{table}
\end{table}
\end{center}

\subsubsection{Generalization to $k$-channels}
We have focused on the 1CK and the 2CK impurities, since there are concrete realizations of these impurities with quantum dots. Yet, it is worthwhile to study the more general $k$-channel Kondo case. In the Fermi liquid fixed points at zero temperature, all the $\psi$-particles are scattered into $\psi$-particles, namely, $|\mathcal{S}_{\psi\psi}|=1$. In the non-Fermi liquid 2CK fixed point, none of the $\psi$-particles are scattered into $\psi$-particles, namely, $|\mathcal{S}_{\psi\psi}|=0$. In the more general $k$-channel Kondo case, however, where $k>1$ channels screen the impurity, a finite part of the $\psi$-particles are elastically scattered off the impurity. At zero temperature, the single $\psi$-particle element of the $\mathcal{S}$-matrix is~\cite{PhysRevB.48.7297}
\begin{equation}
 S^{\rm kCK}_{\psi\psi}=\frac{\cos\left(\frac{2\pi}{2+k}\right)}{\cos\left(\frac{\pi}{2+k}\right)}\;.
\end{equation}
The conductance, up to $\mathcal{O}\left(T/T_K\right)^{\frac{4}{2+k}}$, is~\cite{PhysRevB.48.7297}
\begin{equation}
G_{\rm d}=\frac{e^2}{h}\sin^2(2\alpha)\left[1-\frac{\cos\left(\frac{2\pi}{2+k}\right)}{\cos\left(\frac{\pi}{2+k}\right)}+c_k\left(\frac{T}{T_K}\right)^{\frac{2}{2+k}}\right]\;,
\end{equation}
where the factor $c_K$ can be calculated numerically~\cite{PhysRevB.48.7297}.
The normalized visibility is
\begin{equation}
\eta=\frac{1}{2}\left[1-\frac{\cos\left(\frac{2\pi}{2+k}\right)}{\cos\left(\frac{\pi}{2+k}\right)}+c_k\left(\frac{T}{T_K}\right)^{\frac{2}{2+k}}\right]+\mathcal{O}\left(T/T_K\right)^{\frac{4}{2+k}}\;,
\end{equation}
and since $\mathcal{S}^{\rm kCK}_{\psi\psi}$ is real, the transmission phase is
\begin{equation}
\varphi_t=\frac{\pi}{2}+\mathcal{O}\left(T/T_K\right)^{\frac{4}{2+k}}\;.
\end{equation}
%
%%%%%%%%%%%%%%%%%%%%%%%%%%%%%%%%%%%%%%%%%%%%%%%%%%%%%%%%%%%%%%
%%%%%%%%%%%%%%%%%%%%%%%%%%%%%%%%%%%%%%%%%%%%%%%%%%%%%%%%%%%%%%
%%%%%%%%%%%%%%%%%%%%%%%%%%%%%%%%%%%%%%%%%%%%%%%%%%%%%%%%%%%%%%

\subsection{External dephasing}\label{sec:external_dephasing}
In Sec. 2, we defined the normalized visibility $\eta$, which is the amplitude of the AB oscillations, normalized in a certain way. In Sec. 3, we showed that $\eta$ has a physical meaning, and that it is related to the proportion of the total conductance carried by single-particle scattering. In this subsection we want to comment about the feasibility of $\eta$-measurements.

So far, we have discussed three mechanisms that reduce the normalized visibility: the possibility of non-coherent charge transfer through the dot into many-body states, thermal averaging over a transmission with energy-dependent phase, and averaging over spin-dependent transmission phase. AB oscillations in a real-life experimental setup can also be suppressed by other mechanisms that are not related to the physical properties of the examined impurity.
A real experimental two-path setup is usually coupled to a complicated environment. For example, in an open AB ring setup the shapes of the two paths, the quantum dot(s), the tunnel barriers, and many other components of the setup are all defined by applying voltages to nearby nano-patterned electrodes. Therefore, each component of the system is coupled to an environment (metal electrodes, semiconducting leads) with associated noise and degrees of freedom.

An electron that propagates through the two paths leaves a trace in the environment; equivalently, a propagating electron that interacts with the environment, accumulates a \emph{random} phase~\cite{PhysRevA.41.3436}, $\varphi$. As a result, the amplitude of the AB oscillations is multiplied by the averaged value $\langle e^{i\varphi}\rangle$. The normalized visibility in the presence of the environment is therefore~\cite{imry2002introduction}
\begin{align}
\sqrt{\eta}&=\langle e^{i\varphi}\rangle\frac{\left|\int d\epsilon\left(-\frac{\partial f}{\partial\epsilon}\right)\mathcal{T}_{\psi\psi}\right|}{\sqrt{\int d\epsilon\left(-\frac{\partial f}{\partial\epsilon}\right)2Im\left\{\mathcal{T}_{\psi\psi}\right\}}}\nonumber\\
&\approx \left(1-\frac{1}{2}\langle \delta\varphi^2\rangle\right)\frac{\left|\int d\epsilon\left(-\frac{\partial f}{\partial\epsilon}\right)\mathcal{T}_{\psi\psi}\right|}{\sqrt{\int d\epsilon\left(-\frac{\partial f}{\partial\epsilon}\right)2Im\left\{\mathcal{T}_{\psi\psi}\right\}}}\;.
\end{align}

The details of the coupling to the environment depend on the details of a specific experimental setup. Yet, we can roughly estimate the external dephasing by assuming that the phase-randomness originates mostly from the thermal fluctuations of the environment. At nonzero temperature $T$, the electrodes in the environment suffer from Nyquist noise, and we can estimate $\langle \delta\varphi^2\rangle\sim T$.

Hence, dephasing by the environment can reduce the normalized visibility linearly in the temperature. In the Fermi liquid fixed points, $\eta$ has $T^2$ corrections without external dephasing. This means that at low temperatures the dominant suppression of $\eta$ would be due to external dephasing. In the non-Fermi liquid fixed point of the 2CK, $\eta$ has a $\sqrt{T}$ dependence in the absence of external dephasing. Thus, at low temperatures the change in $\eta$ (enhancement for $\lambda>0$ and reduction for $\lambda<0$), is expected to be stronger than its suppression due to external dephasing.

The relation between the system and the environment is outside the scope of this work. In particular, we do not get into specific models for the environment. We want to note that there are models that treat rigorously the effect of a specific environment on the interference in AB rings (for example, a quantum-point-contact that is coupled to an embedded quantum dot~\cite{0295-5075-39-3-299,PhysRevLett.79.3740}; or a fluctuating magnetic flux~\cite{PhysRevB.65.125315}).

In the 2CK non-Fermi liquid case, a noisy environment can, in principle, turn on relevant operators. Thus, a noisy environment with strong effect on the system would make the observation of the non-Fermi liquid behavior difficult. Hence, if a non-Fermi liquid behavior is indeed observed in an experimental system, it indicates a relatively weak external dephasing.
%%%%%%%%%%%%%%%%%%%%%%%%%%%%%%%%%%%%%
%%%%%%%%%%%%%%%%%%%%%%%%%%%%%%%%%%%%%
\section{Conclusions and discussion}\label{conclusions}
In this work we have focused on information that can be obtained from two-path experiments. Typically, in two-path experiments, the measurable quantity is the transmission phase $\varphi_t$. We showed that the combination of two measurements, the two path conductance together with the conductance of one of the paths (either of the paths), gives us additional physical information about the nature of coherence in the transport.
These two measurements allow us to normalize the amplitude of the flux-dependent conductance, with respect to the independent conductances of the two paths [Eq.~\eqref{eq: definition of eta and phi}]. We showed that the normalized amplitude $\eta$ is related to the fraction of scattering processes that involve only single particles.

We have related $\varphi_t$ and $\eta$ to the single-particle matrix element of the $\mathcal{S}$-matrix.
If there is a linear combination of the two leads (denoted by $\xi$) which is decoupled both from the dot, and from the orthogonal linear combination ($\psi$), then, working in the $\psi-\xi$ basis we showed that $\varphi_t$ and $\eta$ can be used to study $\mathcal{S}_{\psi\psi}$. In the simple case of Fermi liquids at zero temperature, where $\mathcal{S}_{\psi\psi}=e^{2i\delta_\psi}$, $\varphi_t$ turns out to be identical to $\delta_\psi$, and $\eta=1$.

We also showed that in the absence of spin-symmetry, both $\varphi_t$ and $\eta$ are affected by the summation over spin in a standard conductance measurement. At zero temperature, we showed that the phase $\varphi_t$ is locked at $\pi/2$ independent of $\delta_\psi$ [see Eq.~\eqref{eq: phase lock pi/2}], and that $\eta$ is suppressed to below one [see Eq.~\eqref{eq: normalized visibility two spins}]. A proper measurement in this case would involve independent measurement of the transport of each spin.

In the various Fermi liquid fixed points of the Kondo impurities, we have showed that the transmission phase equals the scattering phase shift $\varphi_{ts}=\delta_{\psi s}$. The normalized visibility at zero temperature is $\eta=1$ , and nonzero temperature reduces it with a correction $\sim \left(T/T_K\right)^2$. The small reduction of $\eta$ is due to two different physical effects of the temperature. First, the transmission phase is energy-dependent. When we thermally average over the temperature, $\varphi_t$ remains at its zero-temperature value (to this order of correction), but $\eta$ is reduced. Second, the nonzero temperature allows incoherent scattering processes (the leading irrelevant operator near the fixed point allows the scattering of single-particle states to many-body states). Hence, a small part of the conductance is incoherent and therefore $\eta$ is reduced.

In the non-Fermi liquid fixed point of the 2CK, we find that although there are no single $\psi$-particle to single $\psi$-particle scattering processes, a part of the conductance is still coherent. The transmission phase is $\varphi_t=\pi/2$ despite the fact that $\delta_\psi$ is not defined. The normalized visibility, at zero temperature, is $\eta=1/2$ indicating the fact that exactly half of the conductance is carried by elastic single-particle scattering events~\cite{PhysRevLett.93.107204}. At nonzero temperature, $\eta$ can either be diminished, or be \emph{enhanced} with a $\sim \sqrt{T/T_K}$ behavior. The enhancement is possible since the leading irrelevant operator near the fixed point allows single $\psi$-particle to single $\psi$-particle scattering.

In real experiments, the propagating electrons are subjected to an external dephasing by the environment. We expect a reduction of the normalized visibility due to this external dephasing. Assuming mostly thermal fluctuations in the environment (Nyquist noise), we roughly estimate a linear temperature dependence of the external dephasing. Therefore, near the Fermi liquid fixed points one might not be able to see the predicted $\sim T^2$ reduction of $\eta$. Near the non-Fermi liquid 2CK fixed point, however, the $\sim\sqrt{T}$ dependence is expected to be parametrically stronger than the external dephasing. Thus we expect that measuring this effect would be possible in the presence of external dephasing.

\section*{ACKNOWLEDGMENTS}
We would like to thank Andrew Keller, Oded Zilberberg, Natalie Lezmy, Eran Sela, and Gergely Zarand for useful discussions. This research was supported by the BSF, GIF, the ISF center of excellence program, by the Minerva Foundation, by the Segre Foundation, and by the US NSF under DMR-0906062.
\appendix
\section{Detailed derivation of the connection between the transmission and the $\mathcal{T}$-matrix}\label{s-matrix appendix}
In this appendix, we present in detail the derivation of the relations between the transmission properties (from the left lead through the impurity to the right lead) and the single-particle matrix elements of the $\mathcal{T}$-matrix.

We want to write a scattering matrix that connects incoming states and outgoing states in the leads. In general, these states can be complex many-body states that involve the two leads. A general state in the two leads can be characterized by two numbers, $n_L$ and $n_R$, according to the charge carried in the two leads. There are, of course, many possible states with charges $en_L$ and $en_R$, since states with the same charges in the two leads can differ by multiple particle-hole excitations. We use the notation $|n_L,n_R,i\rangle$ for these states, where the index $i$ labels the possible states with charges $en_L$ and $en_R$.

The scattering matrix, $\mathcal{S}$, connects incoming and outgoing states
\begin{equation}
|n'_L,n'_R,j\rangle_{\rm out}=\mathcal{S}^{n'_L,n'_R,j}_{n_L,n_R,i}|n_L,n_R,i\rangle_{\rm in}\;.\label{eq: s-matrix-definition}
\end{equation}
Charge conservation imposes $n'_L+n'_R=n_L+n_R$, hence, $\mathcal{S}$ is a block-diagonal matrix. We work in the zero-bias limit at low temperature, and therefore only single electrons can be sent from the source and the drain. Hence, we focus only on the block $n_L+n_R=1$ of the $\mathcal{S}$-matrix.

There are two types of states in the subspace of states with $n_L+n_R=1$, single electron states, and many-body states. We can make this distinction since the two leads are free. We denote by $|L\rangle_{\rm in}$ and $|L\rangle_{\rm out}$ the incoming and outgoing single-electron states in the left lead, and similarly $|R\rangle_{\rm in}$ and $|R\rangle_{\rm out}$ in the right lead. We denote the other states, which are many-body states of the form $|1-n,n,i\rangle$, by $|\chi^i_n\rangle$ ($n=0,\pm1,\pm2...$)~. Notice that in the cases $n=0,1$, the states $|\vec{\chi}_{0,1}\rangle$ span only the multi-particle states. The single electron states of the form $|1,0\rangle$ and $|0,1\rangle$ are denoted by $|L\rangle$ and $|R\rangle$~.

Schematically, The $n_L+n_R=1$ block of Eq.~\eqref{eq: s-matrix-definition} is
\begin{equation}
\left(
  \begin{array}{c}
    |L\rangle_{\rm out} \\
    |R\rangle_{\rm out} \\
    |\boldsymbol{\chi}\rangle_{\rm out} \\
  \end{array}
\right)=\left(
  \begin{array}{cccccccc}
    r & t' &  \boldsymbol{A_L}^{\dagger}\\
    t & r' &  \boldsymbol{A_R}^{\dagger}\\
    \boldsymbol{B_L} & \boldsymbol{B_R} & \boldsymbol{C}\\
  \end{array}
\right)
\left(
  \begin{array}{c}
    |L\rangle_{\rm in} \\
    |R\rangle_{\rm in} \\
    |\boldsymbol{\chi}\rangle_{\rm in} \\
  \end{array}
\right)\;,\label{eq:s matrix appendix}
\end{equation}
where the exact definitions for all the terms in \eqref{eq:s matrix appendix} appear in Sec.~\ref{sec: s-matrix discussion} [see Eqs.~\eqref{eq: definition L,R}-\eqref{eq: definition A}]. Here we don't include spin, and generalization of what follows to spinful electrons is straightforward.

Since the $\mathcal{S}$-matrix is unitary and block diagonal, its $n_L+n_R=1$ block is also unitary. This leads to the following relations
\begin{eqnarray}
&&|r|^2+|t|^2+\sum_{n,i} \left|B_L^{ni}\right|^2=1\;,\label{eq:s-unitary1}\\
&&|r'|^2+|t'|^2+\sum_{n,i} \left|B_R^{ni}\right|^2=1\;,\label{eq:s-unitary2}\\
&&|r|^2+|t'|^2+\sum_{n,i} \left|A_L^{ni}\right|^2=1\;,\label{eq:s-unitary3}\\
&&|r'|^2+|t|^2+\sum_{n,i} \left|A_R^{ni}\right|^2=1\;\label{eq:s-unitary4}.
\end{eqnarray}
Consider now the average current at the right lead. As mentioned before, at low temperature and bias voltage we can assume that only single electrons are sent toward the impurity. The average current is
\begin{align}
I&=\frac{e}{h}\int d\epsilon\left[f_l(\epsilon)\left(|t|^2+\sum_{n,i} n\left|B_L^{ni}\right|^2\right)\right.\nonumber\\
&\left.+f_r(\epsilon)\left(|r'|^2+\sum_{n,i} n\left|B_R^{ni}\right|^2-1\right)\right]\nonumber\\
&=\frac{e}{h}\int d\epsilon\left[f_l(\epsilon)\left(|t|^2+\sum_{n,i} n\left|B_L^{ni}\right|^2\right)\right.\nonumber\\
&\left.-f_r(\epsilon)\left(|t'|^2+\sum_{n,i} (1-n)\left|B_R^{ni}\right|^2\right)\right]\label{eq: s-matrix-current1}.
\end{align}
At equilibrium, the current is zero, therefore
\begin{equation}
|t|^2+\sum_{n,i} n\left|B_L^{ni}\right|^2=|t'|^2+\sum_{n,i} (1-n)\left|B_R^{ni}\right|^2\;,\label{eq: equilibrium_relations}
\end{equation}
and the current becomes
\begin{equation}
I=\frac{e}{h}\int d\epsilon\left[f_l(\epsilon)-f_r(\epsilon)\right]\left(|t|^2+\sum_{n,i} n\left|B_L^{ni}\right|^2\right).
\end{equation}
Thus, the conductance is
\begin{equation}
G_{\rm d}=\frac{e^2}{h}\int d\epsilon\left(-\frac{\partial f}{\partial\epsilon}\right)\left(|t|^2+\sum_{n,i} n\left|B_L^{ni}\right|^2\right).\label{eq: Conductance_appendix}
\end{equation}
The proportion of the total conductance carried by coherent single-particle scattering is
\begin{equation}
G_{\rm coh}/G_{\rm d}=\frac{\int d\epsilon\left(-\frac{\partial f}{\partial\epsilon}\right)\left|t\right|^2}{\int d\epsilon\left(-\frac{\partial f}{\partial\epsilon}\right)\left(|t|^2+\sum_{n,i} n\left|B_L^{ni}\right|^2\right)}\;.
\end{equation}

Suppose now, that there is a linear combination of the two leads, $\xi=-\sin(\alpha)L+\cos(\alpha)R$~, which is decoupled both from the impurity and from the orthogonal combination of the leads, $\psi=\cos(\alpha)L+\sin(\alpha)R$~. This is the case, for example, in the Anderson model for a single level quantum dot that is coupled to two leads. The fact that $\xi$ is a free decoupled field simplifies the above expressions as it imposes the following restrictions on the $\mathcal{S}$-matrix in the $\psi-\xi$ basis: $\mathcal{S}_{\xi x}=\mathcal{S}_{x\xi}=0$ ($x=\psi,\vec{\chi}_n$), and $\mathcal{S}_{\xi\xi}=1$. In particular, the restriction $\mathcal{S}_{\psi\xi}=\mathcal{S}_{\xi\psi}=0$ requires $t'=t$ which, together with Eq.~\eqref{eq: equilibrium_relations}, yields the relation
\begin{equation}
\sum_{n,i} n\left|B_L^{ni}\right|^2=\sum_{n,i} (1-n)\left|B_R^{ni}\right|^2\;.\label{eq: equality of transmitted charge}
\end{equation}

Moreover, we can relate $B_L^{ni}$ and $B_R^{ni}$. Since (omitting the in and out subscripts)
\begin{eqnarray}
B_L^{ni}=\langle\chi^i_n|L\rangle=\cos(\alpha)\langle\chi^i_n|\psi\rangle-\sin(\alpha)\langle\chi^i_n|\xi\rangle,&&\\
B_R^{ni}=\langle\chi^i_n|R\rangle=\sin(\alpha)\langle\chi^i_n|\psi\rangle+\cos(\alpha)\langle\chi^i_n|\xi\rangle,&&
\end{eqnarray}
and as $\langle\chi^i_n|\xi\rangle=0$ we get
\begin{eqnarray}
&&B_L^{ni}=\cos(\alpha)\langle\chi^i_n|\psi\rangle\;,\\
&&B_R^{ni}=\sin(\alpha)\langle\chi^i_n|\psi\rangle\;.
\end{eqnarray}
We obtain the relation
\begin{equation}
B_R^{ni}=\tan(\alpha)B_L^{ni}\;.
\end{equation}
Plugging this relation into Eq.~\eqref{eq: equality of transmitted charge} gives
\begin{equation}
(1+\tan^2(\alpha))\sum_{n,i}n\left|B_L^{ni}\right|^2=\tan^2(\alpha)\sum_{n,i}\left|B_L^{ni}\right|^2.
\end{equation}
Thus, we get the important equalities
\begin{eqnarray}
&&\sum_{n,i}n\left|B_L^{ni}\right|^2=\sin^2(\alpha)\sum_{n,i}\left|B_L^{ni}\right|^2,\label{eq: s-matrix sum ruleL}\\
&&\sum_{n,i}n\left|B_R^{ni}\right|^2=\sin^2(\alpha)\sum_{n,i}\left|B_R^{ni}\right|^2\label{eq: s-matrix sum ruleR}.
\end{eqnarray}
Together with Eqs.~(\ref{eq:s-unitary1}) and (\ref{eq:s-unitary2}), the sum rules (\ref{eq: s-matrix sum ruleL}) and (\ref{eq: s-matrix sum ruleR}), tell us that the incoherent part of the conductance, which is carried by single-particle to many-particles scattering processes, can also be determined by the coherent single-particle part of the $\mathcal{S}$-matrix.

Notice also that $\sum_{n,i}\langle\psi|\chi_n^i\rangle\langle\chi_n^i|\psi\rangle=\sum_{ni}|\langle\psi|\chi_n^i\rangle|^2$ is the sum of probabilities to find outgoing states if the incoming state is $|\psi\rangle$. Since we sum over all possible outgoing states besides $|\psi\rangle$ and $|\xi\rangle$, and as $\langle\xi|\psi\rangle=0$ we find that
\begin{equation}
\sum_{n,i}\langle\psi|\chi_n^i\rangle\langle\chi_n^i|\psi\rangle=1-|_{\rm out}\langle\psi|\psi\rangle_{\rm in}|^2=1-|\mathcal{S}_{\psi\psi}|^2,
\end{equation}
so
\begin{eqnarray}
&&\sum_{n,i}\left|B_L^{ni}\right|^2=\cos^2(\alpha)(1-|\mathcal{S}_{\psi\psi}|^2)\;,\\
&&\sum_{n,i}\left|B_R^{ni}\right|^2=\sin^2(\alpha)(1-|\mathcal{S}_{\psi\psi}|^2)\;.
\end{eqnarray}
The conductance~\eqref{eq: Conductance_appendix} can be written as
\begin{equation}
G_{\rm d}=-\frac{e^2}{h}{\int d\epsilon\frac{\partial f}{\partial\epsilon}\left[|t|^2+\sin^2(\alpha)\cos^2(\alpha)(1-|\mathcal{S}_{\psi\psi}|^2)\right]}\;,\label{ea: s-matrix conductance}
\end{equation}
and the contribution of the single-particle processes to the conductance, out of the total conductance is
\begin{equation}
G_{\rm coh}/G_{\rm d}=\frac{\int d\epsilon\frac{\partial f}{\partial\epsilon}\left|t\right|^2}{\int d\epsilon\frac{\partial f}{\partial\epsilon}\left(|t|^2+\sin^2(\alpha)\cos^2(\alpha)(1-|\mathcal{S}_{\psi\psi}|^2)\right)}\;.\label{eq: s-matrix normalized visibility}
\end{equation}

The fact that there is a linear combination of $L$ and $R$ which is decoupled both from the impurity and from the orthogonal linear combination imposes restrictions on the values of $r,t,r',t'$ (since $\mathcal{S}_{\psi\xi}=\mathcal{S}_{\xi\psi}=0$ and $\mathcal{S}_{\xi\xi}=1$). By applying the unitary transformation on the $\mathcal{S}$-matrix one finds that
\begin{equation}
\mathcal{S}_{\psi\psi}=1+\frac{t}{\cos(\alpha)\sin(\alpha)}\;.\label{eq:s-matrix relation between t and s}
\end{equation}
Plugging (\ref{eq:s-matrix relation between t and s}) into \eqref{ea: s-matrix conductance} and \eqref{eq: s-matrix normalized visibility} gives
\begin{align}
&G_{\rm d}=-\frac{e^2}{h}\frac{\sin^2(2\alpha)}{4}\int d\epsilon\frac{\partial f}{\partial\epsilon}\left(\left|\mathcal{S}_{\psi\psi}-1\right|^2+\left[1-|\mathcal{S}_{\psi\psi}|^2\right]\right),\label{eq: s-matrix conductance1}\\
&G_{\rm coh}/G_{\rm d}=\frac{\int d\epsilon\frac{\partial f}{\partial\epsilon}\left|\mathcal{S}_{\psi\psi}-1\right|^2}{\int d\epsilon\frac{\partial f}{\partial\epsilon}\left(\left|\mathcal{S}_{\psi\psi}-1\right|^2+\left[1-|\mathcal{S}_{\psi\psi}|^2\right]\right)}\;.\label{eq: s-matrix normalized visibility1}
\end{align}

At this point we can already see two important features: First, $G_{\rm coh}/G_{\rm d}$ depends only on $\mathcal{S}_{\psi\psi}$ and in particular does not depend directly on $\alpha$. Second, if $|\mathcal{S}_{\psi\psi}|=1$ (but $\mathcal{S}_{\psi\psi}\neq1$) then $G_{\rm coh}/G_{\rm d}=1$, and if $\mathcal{S}_{\psi\psi}=0$ then $G_{\rm coh}/G_{\rm d}=1/2$. In other words, for a zero temperature Fermi liquid theory $\eta=1$, and for a theory where $\psi$ has no single-particle to single-particle scattering processes (like in the 2CK case at zero temperature) $\eta=1/2$.

Using the definition $\mathcal{S}=1+i\mathcal{T}$ for the $\mathcal{T}$-matrix, we can bring \eqref{eq: s-matrix conductance1} and \eqref{eq: s-matrix normalized visibility1} into the form
\begin{align}
&G_{\rm d}=\frac{e^2}{h}\frac{\sin^2(2\alpha)}{4}\int d\epsilon\left(-\frac{\partial f}{\partial\epsilon}\right)2Im\left\{\mathcal{T}_{\psi\psi}\right\}\;,\\
&G_{\rm coh}/G_{\rm d}=\frac{\int d\epsilon\left(-\frac{\partial f}{\partial\epsilon}\right)\left|\mathcal{T}_{\psi\psi}\right|^2}{\int d\epsilon\left(-\frac{\partial f}{\partial\epsilon}\right)2Im\left\{\mathcal{T}_{\psi\psi}\right\}}\;.
\end{align}

\section{Model for a quantum dot impurity embedded into an open AB ring}\label{two paths setup appendix}

\begin{figure}
\includegraphics[bb=156bp 350bp 471bp 511bp,clip,scale=0.9]{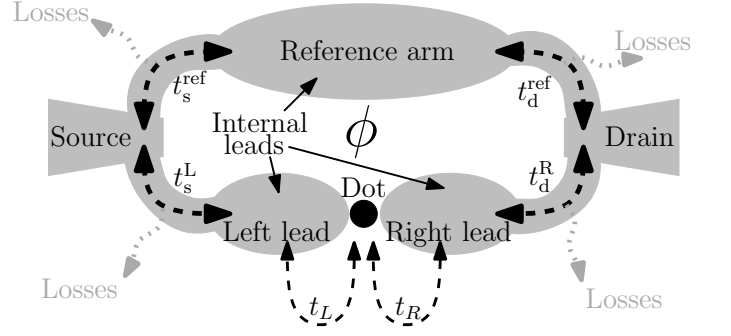}
\caption{Schematic model of a quantum dot embedded into an open AB ring. The four transmission coefficients between the two paths and the external leads ($t_{\rm s}^{\rm ref}$, $t_{\rm d}^{\rm ref}$,$t_{\rm s}^{\rm L}$,$t_{\rm d}^{\rm R}$), are very small. To the lowest order in the external transmission coefficients the propagations along the arms are independent of each other. Because of the losses, time reversal symmetry is broken. We encode the losses in the transmission coefficients.}
\label{fig:system}
\end{figure}

In this appendix, we present a model for a possible setup of a quantum dot that is embedded into an open AB ring. Setups of this kind, can be used to study the transmission through 1CK and 2CK impurities.

Consider the open AB ring setup that is depicted in Fig. \ref{fig:system}. The system contains two external leads (source and drain) and two internal paths. The external leads are coupled to the two paths by four transmission coefficients ($t_s^{\rm ref}$, $t_d^{\rm ref}$, $t_s^{L}$, $t_d^{R}$) which are assumed to be very small. The two possible paths are either through the quantum dot (the lower arm in Fig. \ref{fig:system}) or through the reference arm (the upper arm in Fig. \ref{fig:system}). When an electron is propagating along the lower arm, it has a finite probability to leak outside the system. However, once it gets close enough to the dot we assume that it can only scattered (forward or backward) off the dot. We refer to the area near the dot from the left (right) as left (right) lead (not to be confused with the external leads, source and drain). The Hamiltonian of the system is
\begin{equation}
H=H_{\rm external}+H_{\rm ref}+H_{\rm system}+H_t\;,\label{eq: general H}
\end{equation}
where each of the three first elements on the right hand side of (\ref{eq: general H}) describes one part of the system.
$H_{\rm external}$ describes the external leads
\begin{equation}
H_{\rm external}=\sum_{r=S,D}\sum_{k,s}\epsilon_kc_{rks}^{\dagger}c_{rks}\;,
\end{equation}
where $c_{rks}$ are the annihilation operators of electrons with spin $s$ in external lead $r$.
$H_{\rm ref}$ describes the free electrons in the reference arm. The lower arm is described by the Hamiltonian
\begin{eqnarray}
H_{\rm system}&&=\sum_{i=L,R}\sum_{k,s}\epsilon_kc^{\dagger}_{iks}c_{iks}+H_{\rm dot}\nonumber\\
&&+\sum_{i=L,R}\sum_{ks}\left(t_ic^{\dagger}_{iks}d_s+h.c.\right),\label{eq: lower arm hamiltonian}
\end{eqnarray}
where $c_{rks}$ are the annihilation operators of electrons with spin $s$ in the internal lead $i$, and $d_s$ annihilates an electron with spin
$s$ in the dot. $H_{\rm dot}$ describes the quantum dot itself and any other system that might interact with it but do not interact directly with the other part of the setup (\emph{e.g.} a capacitively coupled gate electrode, other dots etc.). The different parts of the setup are connected via $H_t$
\begin{eqnarray}
H_t&&=\sum_{ks}\sum_{r}t_r^{\rm ref}c_{rks}^{\dagger}c_{{\rm ref},ks}\nonumber\\
&&+\sum_{ks}t_s^{L}c_{Sks}^{\dagger}c_{Lks}+\sum_{ks}t_d^{R}c_{Dks}^{\dagger}c_{Rks}+h.c.\;.
\end{eqnarray}
We don't get into the details of how the setup is coupled to other side leads.

To the lowest order in the external transmission coefficients, $H_t$, the two paths are independent of each other. Therefore, using the definitions of $\varphi_t$ and $\eta$ [see Eq.~\eqref{eq: definition of eta and phi}], the conductance can be written in the form
\begin{equation}
G_{\rm sd}=G_{\rm d}+G_{\rm ref}+2\sqrt{\eta}\sqrt{G_{\rm d}G_{\rm ref}}\cos\left(\frac{e\phi}{\hbar c}+\varphi_t\right),\nonumber
\end{equation}
where $G_{\rm ref}$ is the conductance through the reference arm, and $G_{\rm d}$ is the conductance through the dot.
There is a linear combination of the internal leads, $\xi=-\sin(\alpha)c_L+\cos(\alpha)c_R$, where $\alpha=\arctan(t_R/t_L)$, which is decoupled both from the dot and from the orthogonal combination of the leads, $\psi=\cos(\alpha)c_L+\sin(\alpha)c_R$~. Following the discussion in section \ref{sec: s-matrix discussion} the transmission through the dot is proportional to the $\mathcal{T}$-matrix of the $\psi$-particles, $\mathcal{T}_{s,\psi\psi}$.

So far, we haven't specified what is the Hamiltonian of the dot, $H_{\rm dot}$~. In other words, we haven't specified other systems that interact with the dot (and do not interact directly with the ring). In the following two subsections, we discuss two specific cases: A 1CK case, where the dot is attached to a gate electrode and tuned to form a 1CK impurity, and a 2CK case, where another large dot is coupled to the small dot with appropriate gate electrodes to form a 2CK impurity~\cite{PhysRevLett.90.136602}.

\subsection{Single-channel Kondo}
The dot is capacitively coupled to a gate electrode. If a gate voltage is applied, then at low enough energies, by tuning the
gate voltage and the tunneling barriers between the dot and the ring ($t_{L,R}$),
one can bring the Hamiltonian (\ref{eq: lower arm hamiltonian})
to the form of Kondo Hamiltonian~\cite{PhysRevLett.87.216601}
\begin{align}
H_{\rm system}&=\sum_{k,s}\epsilon_{k}\psi_{ks}^{\dagger}\psi_{ks}+\sum_{k,s}\epsilon_{k}\xi_{ks}^{\dagger}\xi_{ks}\nonumber\\
&+J\sum_{k,s}\sum_{k',s'}\psi_{ks}^{\dagger}\vec{\sigma}_{ss'}\psi_{k's'}\cdot\vec{S}\;,\label{eq: SCK H}
\end{align}
where $\xi=-\sin(\alpha)c_L+\cos(\alpha)c_R$, and $\psi=\cos(\alpha)c_L+\sin(\alpha)c_R$.
$J$ is the Kondo interaction strength, $\vec{\sigma}$ are the three
Pauli matrices, and $\vec{S}$ is the total spin of the dot. Up to second order in $1/T_K$ the $\mathcal{T}_{s, \psi\psi}$-matrix is~\cite{PhysRevB.48.7297}
\begin{equation}
\mathcal{T}_{s,\psi\psi}\left(\epsilon\right)=i\left[2+i\frac{2\epsilon}{T_K}-3\left(\frac{\epsilon}{T_K}\right)^{2}-\left(\frac{\pi T}{T_K}\right)^{2}\right]\;.
\end{equation}

\subsection{Two-channel Kondo}

We can tune the part of the system that is described by $H_{\rm dot}$ to form a 2CK impurity (\emph{e.g.}, by adding another relatively large quantum dot, and couple it to
the small dot~\cite{PhysRevLett.90.136602}). The Hamiltonian (\ref{eq: lower arm hamiltonian}) becomes~\cite{PhysRevLett.90.136602,PhysRevB.69.115316}
\begin{align}
&H_{\rm system}= \sum_{k,s}\epsilon_{k}\psi_{ks}^{\dagger}\psi_{ks}+\sum_{k,s}\epsilon_{k}\xi_{ks}^{\dagger}\xi_{ks}+\sum_{k,s}\epsilon_{k}D_{ks}^{\dagger}D_{ks}\nonumber\\
&+\sum_{k,s}\sum_{k',s'}\left(J_{\psi}\psi_{ks}^{\dagger}\vec{\sigma}_{ss'}\psi_{k's'}+J_{D}D_{ks}^{\dagger}\vec{\sigma}_{ss'}D_{k's'}\right)\cdot\vec{S}\;,
\end{align}
where $D_{ks}$ are the annihilation operators of the large dot, and $J_D$ ($J_{\psi}$) is the strength of the interaction between the spin of the electrons in the large dot (in the $\psi$ lead) and the total spin of the small dot. By
tuning the parameters properly, we can bring the system to
the symmetric point $J_{\psi}\approx J_{D}$, where it displays a
non Fermi liquid behavior~\cite{PhysRevLett.90.136602}. In this case, up to order $1/\sqrt{T_K}$,
the $\mathcal{T}_{s, \psi\psi}$-matrix is~\cite{PhysRevB.48.7297}
\begin{equation}
\mathcal{T}_{s,\psi\psi}\left(\epsilon\right)=i\left(1-3\lambda\sqrt{\pi T}I(\epsilon)\right)\;,\nonumber
\end{equation}
where $I(\epsilon)$ was defined in Eq.~\eqref{eq: definition of I(e)}.


\begin{thebibliography}{76}
\bibitem[{foo({\natexlab{a}})}]{footnote_abstract}
\bibinfo{note}{This is in agreement with the relation between the inelastic
  scattering cross section and the total cross section that was found in
  Refs.~\onlinecite{PhysRevLett.93.107204,PhysRevB.75.235112}}.
\bibitem[{foo({\natexlab{b}})}]{footnote_different_channels}
\bibinfo{note}{In the $k$-channel Kondo case, if there is no charge transfer
  between the different channels, $n_L$ and $n_R$ are the charges in one of the
  channels, and states with the same values of $n_L$ and $n_R$ can also differ
  by excitations in the other $k-1$ channels}.
\end{thebibliography}
\end{document}